\documentclass[prx,twocolumn,superscriptaddress,secnumarabic,balancelastpage,longbibliography]{revtex4-1}
\usepackage{siunitx}
\usepackage{braket}
\usepackage{amssymb}
\usepackage{amsmath,amsthm}
\usepackage[dvipsnames]{xcolor}
\usepackage{chngpage}
\usepackage{lgrind}
\usepackage{color}
\usepackage{graphics}
\usepackage[pdftex]{graphicx}
\usepackage{longtable}
\usepackage{epsf}
\usepackage{bm}
\usepackage{asymptote}
\usepackage{thumbpdf}
\usepackage{verbatim}
\usepackage{url}
\usepackage{MnSymbol}
\usepackage{physics}
\usepackage[colorlinks=true]{hyperref} 
\usepackage{enumitem}

\usepackage{float}
\usepackage{times}
\usepackage{color}
\usepackage{verbatim}
\usepackage{soul,xcolor}
\setstcolor{red}
\usepackage{colortbl}
\usepackage{textcomp}

\usepackage[normalem]{ulem}

\usepackage{tabularx}
\usepackage{bbold}

\makeatletter
\AtBeginDocument{
\g@addto@macro{\appendix}{\renewcommand{\p@subsection}{\@Alph\c@section}}
}
\makeatother

\begin{document}
\newcommand\numberthis{\addtocounter{equation}{1,2}\tag{\theequation}}

\title{Nanoscale magnetometry via collective many-body dynamics in diamond}

\affiliation{Department~of~Physics,~Harvard~University,~Cambridge,~MA~02138,~USA}
\affiliation{Marian Smoluchowski Institute of Physics, Jagiellonian University in Krak\'ow, 30-348 Krak\'ow, Poland} 
\affiliation{Harvard~Quantum~Initiative,~Harvard~University,~Cambridge,~MA~02138,~USA}
\affiliation{Materials~Department,~University~of~California,~Santa~Barbara,~CA~93106,~USA}
\affiliation{Division~of~Engineering~and~Applied~Science,~California~Institute~of~Technology,~Pasadena,~CA 91125,~USA}
\affiliation{Department~of~Physics,~University~of~California,~Santa~Barbara,~CA~93106,~USA}
\affiliation{Department~of~Chemistry~and~Chemical~Biology,~Harvard~University,~Cambridge,~MA~02138,~USA}

\author{Haoyang~Gao$^1$}
\thanks{These authors contributed equally to this work}
\author{Piotr~Put$^{1,2}$}
\thanks{These authors contributed equally to this work}
\author{Nathaniel~T.~Leitao$^1$}
\thanks{These authors contributed equally to this work}
\author{Nazlı~U.~Köylüoğlu$^{1,3}$}
\author{Andrew~Maccabe$^3$}
\author{Mathew~Mammen$^3$}
\author{Siddharth~Dandavate$^1$}
\author{Lillian~B.~Hughes~Wyatt$^{4,5}$}
\author{Leigh~S.~Martin$^1$}
\author{Ania~C.~Bleszynski~Jayich$^6$}
\author{Hongkun~Park$^{1,7}$}
\author{Mikhail~D.~Lukin$^{1,\dagger}$}

\begin{abstract}
Many-body dynamics constitutes a  promising approach for creating correlations between quantum particles which can be used for applications in sensing and metrology~\cite{kitagawa_squeezed_1993}. However,  utilizing this potential for substantial gains in practical settings is a challenging task with  only a very few applications realized to date~\cite{aasi_enhanced_2013,kominis_subfemtotesla_2003}. Here, we demonstrate an approach to nanoscale magnetic sensing~\cite{schirhagl_nitrogen-vacancy_2014,  casola_probing_2018, mamin_nanoscale_2013, hong_nanoscale_2013} enabled by strongly interacting electronic spins  in a room temperature solid. 
By coherently controlling collective many-body dynamics of a dipolar ensemble of $\sim 10^4$ nitrogen–vacancy (NV) centres in diamond with pulsed magnetic field gradients~\cite{put_collective_2025, leitao_scalable_2026}, we demonstrate practical metrological gain up to $7.9(2)\,\mathrm{dB}$ for magnetic signal detection and $8.8(3)\, \mathrm{dB}$ for magnetic noise sensing, fully accounting for experimental overheads. Finally, we combine these methods to demonstrate a momentum-space-resolved sensing modality that enables detection of spatially correlated magnetic noise at continuously tunable length scales down to 50 nanometers. These observations open the door  toward practical applications of interaction-enhanced quantum sensors for nanoscale biological imaging and material characterization.
\end{abstract}

\maketitle

The use of correlated light and matter  to enhance sensing and metrology is currently being explored  across a range of physical platforms. While  substantial reduction of quantum noise and signal enhancement have been demonstrated in a number of  experiments~\cite{franke_quantum-enhanced_2023,pedrozo-penafiel_entanglement_2020,greve_entanglement-enhanced_2022, colombo_time-reversal-based_2022,li_improving_2023, zaporski_quantum-amplified_2025,bornet_scalable_2023,hines_spin_2023,eckner_realizing_2023,aasi_enhanced_2013}, utilizing these techniques for practical tasks is challenging, as most of the state-of-the-art quantum sensing methods achieve the optimal performance in non-interacting regimes~\cite{marshall_high-stability_2025, kim_atomic_2025, zhang_ultrahigh-sensitivity_2023, arunkumar_quantum_2023}. Notable exceptions include  the use of optical squeezing to improve the sensitivity and bandwidth of Laser Interferometer Gravitational-Wave Observatory (LIGO)~\cite{aasi_enhanced_2013} and spin precession synchronization in high density atomic ensembles for spin-exchange relaxation-free (SERF) atomic magnetometry~\cite{kominis_subfemtotesla_2003}.  
 
Atom-like spin defects formed by nitrogen-vacancy (NV) centers in diamond recently emerged as a promising platform for  magnetic sensing,  combining high sensitivity, nanoscale spatial resolution, and compatibility with ambient environments~\cite{schirhagl_nitrogen-vacancy_2014, casola_probing_2018, mamin_nanoscale_2013, hong_nanoscale_2013}. While dipolar interactions present in dense ensembles of NV centers have been generally considered a limitation, as they cause unwanted dephasing that ultimately limits magnetic sensitivity~\cite{zhou_quantum_2020}, it was demonstrated recently that under proper control, these interactions can be  engineered to reduce  quantum projection noise (QPN)~\cite{wu_spin_2025} or, alternatively, to \textit{amplify} the sensing signal~\cite{gao_signal_2025}.
While significant, these proof-of-concept experiments also reveal  the challenges associated with quantum control of solid-state systems, demonstrating approximately 0.5 dB of spin squeezing within sub-ensembles of NV centers~\cite{wu_spin_2025} and reporting  a modest 1.067$\times$ (0.56 dB) signal enhancement~\cite{gao_signal_2025} under idealized conditions.
Here we demonstrate practical interaction-enhanced nanoscale magnetic sensing using a three-dimensional NV ensemble in bulk diamond.
By leveraging a recently developed technique based on pulsed magnetic field gradients that realizes \textit{collective} one-axis-twisting (OAT) dynamics in a positionally disordered dipolar spin ensemble~\cite{put_collective_2025,leitao_scalable_2026}, together with a novel qubit encoding that substantially extends coherence time~\cite{gao_dressed-state_2026}, we achieve a 5.4$\times$ (14.6 dB) signal amplification  (see Fig.~\ref{fig1}d for concept of signal amplification).
We use these interaction-enhanced protocols in several sensing modalities, demonstrating sensitivity gains of up to 7.9 dB for coherent signals and 8.8 dB for noise sensing, fully accounting for the time overhead and other deleterious effects. Finally, we combine the collective OAT dynamics with nanoscale phase encoding via magnetic-field gradients~\cite{arai_fourier_2015} to demonstrate a novel sensing modality that enables momentum-space-resolved detection of spatially correlated magnetic noise with resolution down to $50$~nm (see Fig.~\ref{fig1}e).

\subsection*{Quantum sensor enhanced through magnetic dipole-dipole interactions}

\begin{figure*}
\centering
\includegraphics[width=0.667\linewidth]{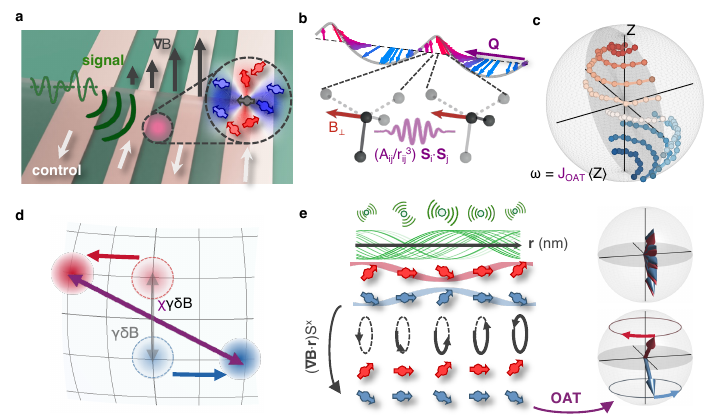}
\caption{\textbf{Interaction-enhanced quantum sensor based on dipolar interactions.} \textbf{a,} Schematic of the experimental system. A diamond hosting a 3D ensemble of interacting NV centers serves as a sensor of external magnetic fields. The diamond is placed on a chip consisting of four parallel microwires that delivers homogeneous MW control. The same wires are diplexed with pulsed electric currents (white arrows) to generate magnetic-field gradients for the preparation of nanoscale spin spirals. Inset: dipolar interactions among NV centers, with colors indicating the angular-dependent sign of coupling. \textbf{b,} Key ingredients enabling interaction-enhanced sensing. Collective OAT dynamics is engineered from dipolar interactions using a controlled nanoscale spin spiral~\cite{put_collective_2025, leitao_scalable_2026}. A bias magnetic field ($B_\perp$) is applied along a direction simultaneously perpendicular to two NV orientations, enabling a dressed-state qubit encoding that substantially improves the coherence time~\cite{gao_dressed-state_2026}. The strength of $B_\perp$ is tuned to realize an $\text{SU}(2)$-symmetric dipolar interaction, maximizing the quality factor of the resulting OAT dynamics. \textbf{c,} Measured polarization dynamics under the collective OAT evolution, for a set of initial states in XZ-plane. The trajectories wrap around the Bloch sphere before significant decay, indicating high-quality collective dynamics. \textbf{d,} Concept of signal amplification used to enhance the sensing performance to global magnetic field. Different magnetic-field signals map to different states on the Bloch sphere, whose separation is amplified by a factor $\chi$ under OAT evolution. The finite extent of the states represents technical noise during readout. 
\textbf{e,} Detecting spatially correlated magnetic noise via collective OAT dynamics and Fourier magnetic imaging~\cite{arai_fourier_2015} (FMI). Top: Noise at wavevector $k$ imprints zero-net-magnetization spin modulations (red and blue denote two noise instances with a $\pi$ phase shift), which cannot be directly amplified by OAT. Bottom: A tailored magnetic-field gradient pulse selectively maps this $k$ mode to $k=0$. This recovers a net $S^z$ polarization for subsequent OAT amplification. We note that initial spin modulations in other $k^\prime$ modes are not mapped to $k=0$ by this gradient pulse, thereby providing momentum-space resolved detection of magnetic noise.
}
\label{fig1}
\end{figure*}

Our experimental platform comprises a three-dimensional (3D), high-density ($\sim$0.5~ppm) ensemble of electronic spins associated with nitrogen–vacancy (NV) centers in diamond (Fig.~\ref{fig1}a), coupled via magnetic dipole-dipole interactions. The diamond is placed on a chip consisting of four microwires (Fig.~\ref{fig1}a) that deliver homogeneous microwave (MW) control across the entire ensemble. The same wires are diplexed with pulsed electric currents of independently programmable magnitude and polarity, enabling the generation of strong, time-dependent magnetic field gradients (Methods). These gradients are used to prepare spiral-like nanoscale spin textures (Fig.~\ref{fig1}b, top), which enable sub-diffraction-limited spatial encoding~\cite{arai_fourier_2015} and provide the mechanism for controlling dipolar interactions for enhanced sensing performance~\cite{put_collective_2025, leitao_scalable_2026}.

\begin{figure*}
\centering
\includegraphics[width=\linewidth]{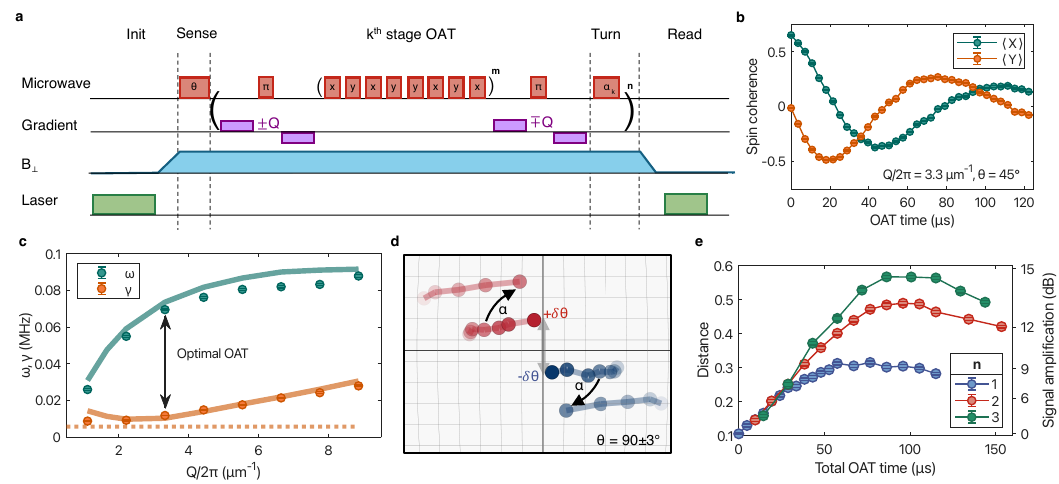}
\caption{\textbf{Signal amplification through collective many-body dynamics.}  
\textbf{a,} Experimental sequence realizing signal amplification. The spin ensemble is optically polarized at zero magnetic field, after which the perpendicular field $B_{\perp}$ is ramped to $362.4~\mathrm{G}$ to tune the dipolar interaction to the form of Eq.~(\ref{eq:Heisenberg_interaction}). A MW pulse of angle $\theta=\pi/2 \pm \delta\theta$ mimics the sensing signal, followed by an OAT stage that amplifies the encoded response. The OAT dynamics are implemented through pulsed control of magnetic-field gradients and MW drive~\cite{put_collective_2025}: a conical spin spiral is wound, evolved under a generalized (XY8)$_m$ sequence that decouples on-site disorder, and then unwound at the end of the quench. An optional MW turn pulse $\alpha_k$ enables multi-stage amplification, for which the ``OAT+turn" sequence is repeated $n$ times. The OAT-generating wavevector is alternated between $+Q$ and $-Q$ in odd/even amplification stages, for suppressing decoherence (see Extended Data Fig.~\ref{ED_fig5} and Supplement). The global spin state is finally read out by ramping $B_{\perp}$ back to zero and measuring the fluorescence under laser illumination. In practice, the ramping of $B_{\perp}$ is achieved by turning on an auxiliary pulsed magnetic field during initialization and readout (Methods), which temporarily cancels the bias field $B_{\perp}$.
\textbf{b,} Measured polarization dynamics at polar angle $\theta = 45^{\circ}$, demonstrating high-quality collective OAT dynamics.  
\textbf{c,} Measured twisting rate $\omega$ and transverse decay rate $\gamma$ as a function of spiral wavevector at polar angle $\theta = 45^{\circ}$. Solid curves show predictions from the system model (Methods).
\textbf{d,} Measured dynamics for a pair of initial states with sensing angle $\theta = 90\pm3^{\circ}$, demonstrating amplification of a small signal. Black arrows indicate the turn pulse used for multi-stage amplification, after which faster twisting dynamics lead to larger amplification.
\textbf{e,} Measured distance (in units of the Bloch-sphere radius) between the initial-state pair in d as a function of total many-body evolution time, for single-stage ($n=1$) and multi-stage ($n=2,3$) protocols. The point at $t=0$ corresponds to the initial distance $2\mathrm{sin}\left(3^\circ\right)\approx 0.105$. Errors represent 1~s.d. accounting statistical uncertainties. 
}
\label{fig2}
\end{figure*}

To achieve substantial metrological gain from dipolar interaction, we employ a collective, OAT-like dynamics generated by the evolution of conically wound nanoscale spin spirals (Fig.~\ref{fig1}b, top), as introduced in Refs.~\cite{put_collective_2025, leitao_scalable_2026}. This dynamics manifests as a Z-dependent precession of the collective Bloch sphere (Fig.~\ref{fig1}c) — where $Z\equiv\frac{2}{N}\sum_iS_i^z$ represents the normalized collective spin — producing a shear that increases the separation between nearby points on the Bloch sphere (Fig.~\ref{fig1}d) and thereby amplifies the sensor’s response to an external signal. A key advance relative to previous work~\cite{put_collective_2025} is a substantial extension of the coherence time of the many-body evolution, achieved using a dressed-state qubit encoding realized by applying a bias magnetic field ($B_\perp$) perpendicular to the NV crystallographic axes~\cite{gao_dressed-state_2026}. The direction of $B_\perp$ is chosen to be simultaneously perpendicular to two distinct NV orientations (Fig.~\ref{fig1}b, bottom), effectively doubling both the density of participating spins and the strength of their mutual interactions.

Specifically, by fixing  $B_\perp$  at $362.4~\mathrm{G}$, the dipolar interaction is tuned from a generic XXZ Hamiltonian to a $\text{SU}(2)$-symmetric form~\cite{gao_dressed-state_2026},
\begin{equation}
    H^{\text{SU}(2)}_{\mathrm{dd}} = -\sum_{ij} \, \frac{8}{3}\frac{J_0}{r_{ij}^3} A_{\hat{\eta}}(\hat{r}_{ij}) \, \vec{S}_i \cdot \vec{S}_j ,
\label{eq:Heisenberg_interaction}
\end{equation}
where $J_0\equiv \left(2\pi\right)\times 52~\mathrm{MHz}\cdot\mathrm{nm}^3$ sets the dipolar interaction scale, $r_{ij}$ denotes the separation between NV centers, and $A_{\hat{\eta}}(\hat{r})= \left(3(\hat{\eta}\cdot\hat{r})^2-1\right)/2$ captures the angular dependence of the dipolar interaction, with quantization axis $\hat{\eta}\equiv\hat{B}_\perp$ determined by the direction of the perpendicular field.
This enhanced symmetry maximizes the quality factor of the collective spin dynamics~\cite{put_collective_2025, leitao_scalable_2026}, enabling the high-quality OAT evolution shown in Fig.~\ref{fig1}c.

To realize the signal amplification illustrated in Fig.~\ref{fig1}d, we follow the protocol illustrated in Fig.~\ref{fig2}a. We begin by optically polarizing the spin ensemble ($\sim$8000 spins) at zero magnetic field, where the optical pumping mechanism is not perturbed by the perpendicular field~\cite{tetienne_magnetic-field-dependent_2012,doherty_nitrogen-vacancy_2013}. We then adiabatically ramp $B_\perp$ to 362.4~G, thereby tuning the dipolar interaction to the form of Eq.~(\ref{eq:Heisenberg_interaction}). The sensing step is mimicked by applying a MW pulse with rotation angle $\theta=\pi/2 \pm \delta\theta$, where $\delta\theta \ll 1$. This is follow by the amplification step, during which the system undergoes $n$ stages of OAT dynamics, interleaved with global MW turn pulses to further enhance the amplification beyond single-stage evolution (as discussed later). Finally, $B_\perp$ is ramped back to zero to perform global spin readout, which relies on the same optical mechanism as the initialization. The OAT dynamics in this protocol is implemented as a three-step procedure including spin spiral winding, many-body quench, and spiral unwinding (see Methods and Refs.~\cite{put_collective_2025, leitao_scalable_2026}).

High-quality OAT dynamics are essential for achieving a large metrological gain. We benchmark these dynamics by preparing an initial state tilted by $\theta=45^\circ$ in the XZ-plane and measuring its X and Y polarization as a function of evolution time (Fig.~\ref{fig2}b). We observe full oscillations corresponding to complete wrapping around the Bloch sphere, indicating high-quality collective OAT dynamics. Repeating this measurement for multiple spiral wavevectors $Q$, we extract both the early-time twisting rate $\omega$ and the decay rate $\gamma$ of the transverse (XY) coherence (Fig.~\ref{fig2}c). We observe a twisting rate on the scale of the dipolar interaction ($\sim 0.08~\mathrm{MHz}$), and a clear separation of timescales between twisting and decay, with the quality factor $\omega/\gamma$ substantially enhanced by the perpendicular-field dressed-state encoding compared to previous work (Ref.~\cite{put_collective_2025}, see Extended Data Fig.~\ref{ED_fig3} for explicit comparisons). From these measurements, we identify the optimal spiral wavevector that maximizes the quality factor, $Q/2\pi = 3.3~$\textmu$\mathrm{m}^{-1}$, which is used throughout this work.

We next perform signal amplification experiments, beginning with single-stage ($n=1$) OAT evolution. We prepare a pair of initial states that mimics a weak sensing signal, corresponding to $\theta = 90^\circ \pm 3^\circ$, and track their polarization dynamics on the Bloch sphere (Fig.~\ref{fig2}d). We observe a shear dynamics consistent with Fig.~\ref{fig1}d. 
The Bloch sphere distance between the pair is plotted in Fig.~\ref{fig2}e (blue curve), yielding an amplification factor of $3.00(7)\times$ at a quench time of approximately 75~\textmu s.

To further enhance signal amplification, we employ a multi-stage amplification strategy. After the first OAT stage, a MW turn pulse is applied to rotate the amplified Bloch-sphere separation — which lies predominantly along the Y direction — toward the Z axis (Fig.~\ref{fig2}d, black arrows). This increased separation along Z leads to faster amplification during the subsequent OAT stage due to the \textit{Z-dependent} precession rate. In Fig.~\ref{fig2}d, this manifests as a larger separation between neighboring points following the turn pulse, even though the time intervals between points are identical to those before the pulse. The Bloch-sphere distances corresponding to two- and three-stage amplification ($n=2,3$) are shown in Fig.~\ref{fig2}e (red and green curves), yielding amplification factors of $4.7(1)\times$ and $5.4(1)\times$, respectively. These values correspond to metrological gains of $13.4(2)\,\mathrm{dB}$ and $14.6(2)\,\mathrm{dB}$, for the task of sensing an instantaneous rotation. The achieved gain substantially exceeds previously reported values ($0.56(5)~\mathrm{dB}$~\cite{gao_signal_2025}) and represents an upper bound on the enhancement achievable in practical sensing applications, where additional polarization loss and time overhead must be taken into account.

\subsection*{Interaction-enhanced sensing of global AC magnetic field}

\begin{figure*}
\centering
\includegraphics[width=0.667\linewidth]{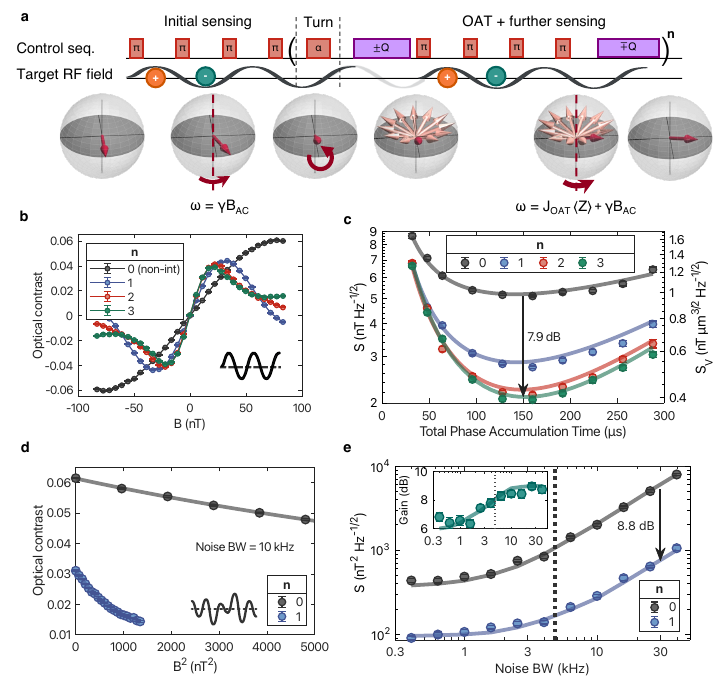}
\caption{\textbf{Interaction-enhanced sensing of global magnetic signal and noise.}  
\textbf{a,} Experimental protocol for interaction-enhanced sensing of an AC magnetic field. An initial sensing stage consists of a train of $\pi$-pulses synchronized to the target radio frequency (RF) signal, acquiring a phase that is subsequently rotated toward the Z axis by a turn pulse $\alpha$. The resulting Z polarization is then amplified by an OAT stage, during which the disorder-decoupling $\pi$-pulses are frequency-synchronized to the target signal and phase-locked to the initial sensing stage, enabling further sensing while amplifying previously accumulated signal. Analogous to Fig.~\ref{fig2}a, this stage can be repeated multiple times for progressively faster amplification. Bloch-sphere representations of the spin state at different stages are shown below the sequence.
\textbf{b,} Sensor response to the amplitude of an applied coherent magnetic signal, showing a steeper response for interaction-enhanced protocols ($n=1,2,3$) compared with the non-interacting case ($n=0$). Data are acquired at a fixed total phase-accumulation time of 160~\textmu s.
\textbf{c,} Sensitivity to coherent signal for non-interacting and interaction-enhanced protocols, demonstrating 7.9(2)~dB metrological gain at the optimum. Solid curves show predictions from a phenomenological model (Methods). 
\textbf{d,} Sensor response to the power of an applied magnetic noise with a bandwidth of $10~\mathrm{kHz}$, showing a steeper initial response in the interaction-enhanced case ($n=1$). Data are acquired at numerically optimized phase-accumulation times (Methods).
\textbf{e,} Sensitivity to noise power as a function of noise bandwidth, demonstrating up to $8.8(3)~\mathrm{dB}$ metrological gain (inset). Dashed vertical line indicates the sensor coherence time $1/T_2$. Solid curves show predictions from the same phenomenological model as in c. Errors represent 1~s.d. accounting statistical uncertainties.
}
\label{fig3}
\end{figure*}

We next apply this signal amplification strategy to the detection of weak, global magnetic fields. We focus on AC magnetic fields in the $\sim$MHz range, which are conventionally detected using modified Ramsey sequences in which a train of $\pi$-pulses is synchronized to the frequency of the target field, $f = 1/(2\tau)$~\cite{taylor_high-sensitivity_2008, degen_quantum_2017}, where $\tau$ is the interpulse spacing.

Adapting interaction-enhanced protocols to this setting requires considerations beyond the conventional  approach of initial sensing followed by a separate amplification stage. In particular, at the optimal phase-accumulation time for conventional sensing — typically comparable to the sensor coherence time $T_2$~\cite{degen_quantum_2017} — the spin ensemble has already experienced substantial polarization loss. This reduces the effective twisting rate in subsequent OAT dynamics and limits the achievable amplification. Moreover, a separate amplification stage introduces additional time overhead, further degrading the practical metrological gain.

We mitigate these limitations by implementing a protocol that enables simultaneous sensing and amplification, thereby reducing the effective time overhead to near zero. This approach exploits the fact that sensing of a global magnetic field commutes with all stages of the spiral-based OAT and therefore can be done simultaneously.
Specifically, we employ the protocol in Fig.~\ref{fig3}a, which contains an initial sensing stage, a global MW turn pulse, and an OAT stage. The initial sensing stage contains a $\pi$-pulse train synchronized to the target magnetic signal frequency, and the accumulated phase is rotated toward the Z axis by the turn pulse, such that it can be amplified by subsequent OAT. During the OAT stage, the disorder-decoupling $\pi$-pulses are frequency-synchronized to the target signal and phase-locked to the initial sensing stage, enabling coherent amplification of previously accumulated phase \emph{while further phase is continuously acquired}. 
Analogous to the pure amplification protocol, this combined sensing–amplification stage can be repeated $n$ times, with appropriately chosen turn angles, to achieve progressively faster amplification while performing continuous sensing. The durations of the initial sensing and OAT stages, as well as the turn angles, are numerically optimized (Methods).

We implement this protocol and measure the spin coherence along the optimal readout axis (sine magnetometry~\cite{degen_quantum_2017}) as a function of the amplitude of a coherent, calibrated magnetic field, for a fixed total phase-accumulation time (Fig.~\ref{fig3}b). Near zero field, the magnetic response is approximately linear, with interaction-enhanced protocols ($n=1,2,3$) exhibiting substantially steeper slopes than the non-interacting case ($n=0$), at the cost of reduced dynamic range. From this response, we extract the magnetic sensitivity $\mathrm{S}$ as a function of total phase-accumulation time (Fig.~\ref{fig3}c), with full consideration of the overhead of the interacting protocol (Methods).
We observe superior sensitivity for interaction-enhanced sensing, reaching a metrological gain of $7.9(2)\,\mathrm{dB}$ for the three-stage ($n=3$) protocol under optimal phase-accumulation time, relative to the non-interacting sensor. The resulting volume-normalized magnetic sensitivity is among the best reported to date~\cite{arunkumar_quantum_2023,masuyama_extending_2018,wolf_subpicotesla_2015,zhou_robust_2023,gao_dressed-state_2026}, demonstrating a practical interaction-enabled enhancement in a state-of-the-art nanoscale sensor.

We similarly apply interaction-enhanced sensing to the detection of incoherent magnetic noise, relevant for nano- to microscale NMR based on statistical polarization~\cite{lovchinsky_nuclear_2016, bucher_quantum_2019}, and for condensed-matter applications where control over the target sample may be limited~\cite{rovny_nanoscale_2024}. We benchmark this regime by measuring the response of $\langle X\rangle$ spin coherence (cosine magnetometry~\cite{degen_quantum_2017}) to a noisy magnetic field centered at $f=1/\left(2\tau\right)$, for both non-interacting ($n=0$) and single-stage amplified ($n=1$) sensors (Fig.~\ref{fig3}d). Despite reduced initial spin coherence, the interacting sensor exhibits a steeper response to noise power. 

We then sweep the bandwidth of the applied magnetic noise and compare the magnetic sensitivity of non-interacting ($n=0$) and interaction-enhanced ($n=1$) sensors across noise regimes (Fig.~\ref{fig3}e). Interaction-enhanced sensing consistently outperforms the non-interacting case, from narrow-band (slowly varying) noise to broadband (Markovian) noise. 
The metrological gain increases with noise bandwidth (inset of Fig.~\ref{fig3}e), reaching $8.8(3)\,\mathrm{dB}$ in the Markovian regime, in good agreement with theoretical predictions. In this limit, phase accumulation transitions from ballistic to diffusive, making it more efficient to allocate the finite coherence time to amplification rather than simply extending the phase-accumulation time, resulting in larger gain. Accounting for the $\sqrt{t}$ scaling of signal-to-noise ratio (SNR) to the total integration time in incoherent noise sensing~\cite{degen_quantum_2017}, the observed gain corresponds to an approximately $57\times$ speedup of magnetic-noise detection.

\subsection*{Momentum-space-resolved sensing of spatially correlated magnetic noise}

While single-point sensing is the most common operating mode of NV-based magnetometry, recent experiments have demonstrated the ability to probe \emph{spatial correlations} of magnetic noise on sub-micrometer length scales~\cite{rovny_nanoscale_2022,le_wideband_2025, ji_correlated_2024, cambria_scalable_2025}. In these approaches, a pair (or a set of pairs) of spins is interrogated individually, and the covariance between their readouts is used to infer magnetic-field correlations at their fixed pairwise separation. Complementary to this approach that relies on the coincident presence of spin pairs at the length scale of interest, we now show that collective OAT dynamics can be harnessed to probe magnetic correlations at continuously tunable length scales through detection-noise-robust measurements of global $\langle S^x\rangle$ polarization~\cite{rovny_multi-qubit_2025}.

\begin{figure}
\centering
\includegraphics[width=\columnwidth]{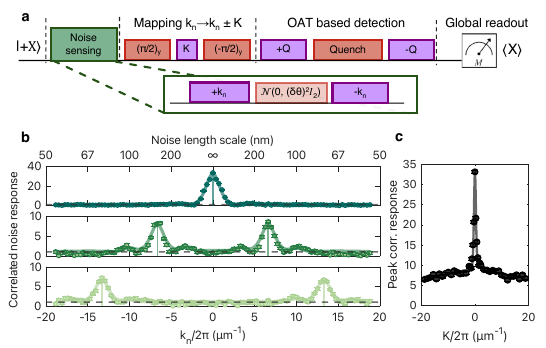}
\caption{\textbf{Interaction-enabled correlation sensing at nanometer length scales.}
\textbf{a,} Protocol for interaction-enabled correlation sensing, where purple (red) squares represent pulses of magnetic-field gradient (MW drive). The protocol starts with a noise sensing stage which encodes the spatially correlated magnetic-field noise into $S^z$ modulation. In the subsequent mapping stage, spin modulation components at wavevector $k_n$ are shifted to $k_n \pm K$ by gradient-induced winding along X, implemented as a Z-winding surrounded by a pair of $\pi/2$ pulses. The resulting spin state then undergoes an OAT stage, which selectively amplifies globally correlated noise (\textit{i.e.} noise around $k_n \pm K=0$), enabling momentum-space-resolved sensitivity. Inset shows the protocol for generating synthetic correlated noise in the experiment. A stochastic MW pulse is conjugated by a pair of spiral winding and unwinding operations, mimicking spatially correlated magnetic noise at wavevector $k_n$.
\textbf{b,} Measured momentum-space response function $F(k)$ defined in Eq.~(\ref{eq:response_function}). The three panels correspond to different values of $K$, indicated by corresponding vertical lines. The solid curves are fits to a mean-field spin-wave model (see Supplement). The horizontal dashed lines indicate the response of a non-interacting sensor, for which $F(k)\equiv 1$ for all $k$. \textbf{c,} Measured peak response along $k_n=K$, demonstrating sensitivity down to $\sim 50~$nm. Errors represent 1~s.d. accounting statistical uncertainties.}
\label{fig4}
\end{figure}

Our approach exploits the fact that \emph{collective} OAT dynamics selectively amplify \emph{globally correlated} magnetic noise (\textit{i.e.} noise at $k_n=0$, see Supplement Sec.~III for justification).
To quantitatively characterize this $k$-selectivity, we introduce a momentum-space response function $F(k)$. Motivated by non-interacting cosine magnetometry, where a phase fluctuation with standard deviation $\delta\theta$ reduces the Bloch-vector length by a factor $\mathrm{e}^{-\left(\delta\theta\right)^2/2}$, we define $F(k)$ through
\begin{equation}
    \langle S^x\rangle_\mathrm{noisy} = \langle S^x\rangle_\mathrm{noiseless}\mathrm{e}^{-F(k)\left(\delta\theta\right)^2/2},
\label{eq:response_function}
\end{equation}
where the subscripts denote measurements with and without magnetic noise. For a non-interacting sensor, this definition yields $F(k)\equiv 1$ for all $k$.

We experimentally measure $F(k)$ using the protocol shown in Fig.~\ref{fig4}a, initially omitting the ``mapping" stage. 
The protocol consists of a sensing stage and a subsequent OAT stage, with the latter intended to amplify the sensor's response around noise wavevector $k_n=0$. To mimic sensing of a spatially correlated magnetic noise at wavevector $k_n$, we artificially imprint a stochastic Z-modulation at this wavevector. This is achieved by conjugating a stochastic global MW ``noise pulse" with a pair of spiral winding and unwinding operations (Fig.~\ref{fig4}a, inset), where the noise pulse is drawn from a Gaussian distribution $\mathcal{N}(0,(\delta\theta)^2 I_2)$ with random X and Y quadratures.
Extracting $F(k)$ according to Eq.~(\ref{eq:response_function}), we observe a sharply peaked response centered at $k_n = 0$ (Fig.~\ref{fig4}b, top panel), with a width set by the finite system size (see Supplement). This sharply selective response enables the detection of weak correlated noise even in the presence of a strong uncorrelated background.

To extend this sensing capability beyond $k=0$, we introduce a mapping stage (Fig.~\ref{fig4}a), implemented as a gradient-induced winding along the X direction. This operation maps a Z-modulation at wavevector $k_n$ to $k_n \pm K$ (see Fig.~\ref{fig1}e for the case of $K=k_n$), effectively shifting the peak of the response function from $k_n=0$ to $k_n=\pm K$. This behavior is demonstrated in Fig.~\ref{fig4}b (bottom two panels) for two different values of $K$.
The reduced peak amplitude arises because the mapping stage transfers approximately half of the noise power into Y-modulation, which also gives rise to the smaller side peaks adjacent to the main response (see Supplement). 
We also measure the peak response along $k_n=K$ while sweeping $K$ (Fig.~\ref{fig4}c), demonstrating sensitivity to correlated noise down to $\sim 50~$nm length scale.
This mapping capability, combined with the sharply peaked sensor response, enables programmable access to magnetic noise correlations at tunable nanometer length scales.

\subsection*{Discussion and outlook}
Our experiments demonstrate \emph{practical metrological gain} enabled by dipole–dipole interactions in NV-ensemble-based magnetometer, overcoming the challenges posed by dipolar anisotropy and positional disorder~\cite{wu_spin_2025,gao_signal_2025}.
This method of sensitivity enhancement has direct implications for applications ranging from biological imaging to materials characterization, enabling the detection of signals that would otherwise remain inaccessible. 
Moreover, we realize a \emph{new sensing modality}: momentum-space-resolved detection of correlated magnetic noise. This capability is particularly relevant for condensed-matter systems~\cite{rovny_nanoscale_2024}, where periodic structures like Wigner crystals~\cite{li_imaging_2021} and ground-state spiral-orders~\cite{lebech_magnetic_1989,haykal_antiferromagnetic_2020} are long-standing and actively studied topics.
The demonstrated tunability of the sensing length scale through gradient-based mapping, combined with frequency selectivity set by the interpulse spacing, enables convenient access to magnetic excitations that are naturally described in the $(k,\omega)$ basis at radio (and microwave) frequencies through correlated $T_2$ ($T_1$) spectroscopy~\cite{rovny_nanoscale_2024, koyluoglu_interaction-enhanced_2026}. 

Looking ahead, two research directions offer complementary opportunities for further metrological gain in interaction-enhanced quantum sensing. First, extension of the spin ensemble coherence time — achieved via cryogenic operation or faster dynamical decoupling and microwave pulse engineering — can probe the intrinsic collective limit of the spin-transport-based OAT mechanism~\cite{leitao_scalable_2026}. Second, sensitivity can be improved by integrating our approach with improved spin readout. In the present work, photon collection efficiency through a thick diamond substrate constitutes a key limitation~\cite{nishimura_investigations_2025,siyushev_monolithic_2010}.
This can be alleviated by employing thinner samples in the form of diamond membranes~\cite{guo_direct-bonded_2024}, or by integration of nanophotonic structures~\cite{li_efficient_2015,huang_monolithic_2019} and the use of custom-designed, high-efficiency microscope objectives~\cite{xu_high-efficiency_2019}.
Beyond photon collection, techniques such as spin-to-charge conversion~\cite{shields_efficient_2015} (SCC), resonant optical readout (RR) at cryogenic temperatures~\cite{robledo_high-fidelity_2011,irber_robust_2021}, and repetitive readout using a nuclear-spin memory~\cite{arunkumar_quantum_2023} provide viable pathways to higher readout fidelity. 
Ultimately, if readout fidelity can be improved to the level where QPN is resolved~\cite{maier_readout_2026}, collective OAT dynamics could enable operation beyond SQL through the generation of spin-squeezed states~\cite{kitagawa_squeezed_1993, leitao_scalable_2026} or time-reversed amplification protocols~\cite{davis_approaching_2016,gao_signal_2025,leitao_optimally_2026}, realizing practical quantum advantage in nanoscale sensing.

\clearpage
\newpage

\section*{Methods}

\noindent\textbf{Diamond sample} \\
The diamond sample used in this work is the same bulk crystal as in Refs.~\cite{put_collective_2025,gao_dressed-state_2026}. 

Diamond homoepitaxial growth and nitrogen doping were performed via plasma-enhanced chemical vapor deposition (PECVD) using a SEKI SDS6300 reactor on a (100)-oriented electronic-grade substrate (Element Six Ltd.). The substrate was fine-polished to a surface roughness of 200--300~pm and etched by 4--5~\textmu m to remove polishing-induced strain. Growth used a 750~W plasma with 0.5\% $^{12}$CH$_4$ in 400~sccm H$_2$ at 25~torr and $\sim$730~$^\circ$C. This process produced a 125~nm isotopically purified (99.998$\%$ $^{12}$C) buffer layer, followed by a 185~nm $^{15}$N-doped layer (1~sccm $^{15}$N$_2$), and a 100~nm $^{12}$C capping layer. Secondary ion mass spectrometry (SIMS) measurements confirmed isotopic purity and layer thicknesses.

NV centers were subsequently generated by electron irradiation using a 200~keV transmission electron microscope (ThermoFisher Talos F200X~G2), producing spots with doses from $10^{17}$--$10^{21}$~e$^-$/cm$^2$. All experiments were conducted at a spot with irradiation dose $2.4\times10^{19}$~e$^-$/cm$^2$. The sample was annealed at 850~$^\circ$C for 6~h in Ar/H$_2$ to form NV centers, cleaned in boiling triacid (H$_2$SO$_4$:HNO$_3$:HClO$_4$ = 1:1:1), and oxygen-terminated at 450~$^\circ$C in air to stabilize the NV$^-$ charge state.

The density of NV centres was estimated from XY16 decay times under on-axis magnetic-field alignment. For the two NV orientation groups used in this work, the measured $1/\mathrm{e}$ decay times are $20.6~$\textmu s and $19.9~$\textmu s, respectively, corresponding to NV$^-$ densities of 257~ppb and 266~ppb. The conversion between XY16 decay time and NV$^-$ density is based on numerical simulations assuming dipolar-interaction-dominated decay.
\\

\noindent\textbf{Circuit designs}\\
A high-level schematic of the control circuitry is shown in Extended Data Fig.~\ref{ED_fig1}. The system comprises three custom-built components: a gradient chip, a current pulser, and an interfacing printed circuit board (PCB).

The gradient chip (Extended Data Fig.~\ref{ED_fig2}a) is diplexed to deliver three control fields: the MW drive, the magnetic-field gradient, and a pulsed perpendicular magnetic field that temporarily cancels the static $B_\perp$ during spin initialization and readout~\cite{gao_dressed-state_2026}, enabling the field ramp shown in Fig.~\ref{fig2}a. We adopt this field-cancellation strategy rather than pulsing $B_\perp$ directly to improve field homogeneity during many-body dynamics and to reduce the duty cycle associated with resistive heating. The chip consists of four gold microwires defined lithographically on the surface of a polycrystalline diamond substrate (Applied Diamonds, optical grade, thickness 130~\textmu m), chosen for its high thermal conductivity and optical transparency. Near the center of the chip, the four wires taper to widths of 10, 5, 5, and 10~\textmu m, respectively, and are separated by equal gaps of 6~\textmu m. The NV ensemble is located between the central two wires, $\sim 3$~\textmu m above them. The current polarities used to generate the three control fields are indicated in Extended Data Fig.~\ref{ED_fig2}b-d and are chosen to optimize the homogeneity of the MW drive, the strength and linearity of the gradient, and the magnitude of the pulsed perpendicular field.

To interface external control signals with the chip, we employ a custom PCB (Extended Data Fig.~\ref{ED_fig2}e,f), which independently routes signals to each of the four microwires. MW signals are delivered through SMPM connectors. During typical operation, two connectors on one side of the PCB are terminated to avoid reflections, while two MW signals with a relative phase shift of $180^\circ$ (obtained by splitting a single MW signal through a Fairview Microwave FMCP1155 SMA 180-degree hybrid coupler) are fed into the remaining two ports. This configuration produces a homogeneous MW drive at the location of the spin ensemble.

The pulsed magnetic-field gradient and perpendicular field are produced using a custom current pulser (Extended Data Fig.~\ref{ED_fig2}g), enabling short ($>15$~ns) pulses with switchable polarity and an amplitude up to $\sim 1.1~\mathrm{A}$. The pulser is the same design as reported in Ref.~\cite{put_collective_2025}, choosing the version with power operational amplifiers (OPA564) instead of voltage regulators. This version allows the output currents to dynamically track the reference voltages, enabling fast ($<5~$\textmu s) switching between the gradient and perpendicular-field configurations. The output currents of these two configurations are controlled by two sets of four reference voltages, which are externally selected using a quad single-pole double-throw (SPDT) switch (Analog Devices ADG1534). The reference voltages are supplied by analog output channels of data acquisition (DAQ) cards.
\\

\noindent\textbf{Sensing signal generation} \\
The target RF magnetic field used in Fig.~\ref{fig3}b,c is generated by an external coil positioned approximately $2~\mathrm{cm}$ from the diamond sample. The coil is driven by an analog output channel of an arbitrary waveform generator (AWG; Tektronix 7122C), ensuring phase locking with the $\pi$-pulse train used for sensing. The AWG output is attenuated by $15~\mathrm{dB}$ before driving the coil to match the dynamic range of the NV magnetometer.

The RF magnetic noise used in Fig.~\ref{fig3}d,e is generated by the same coil, driven by a signal generator (Agilent 33522A). The signal generator outputs a sine wave with its amplitude modulated by an internal noise source, producing RF noise centered at $f=12.5~\mathrm{MHz}$.
\\

\noindent\textbf{OAT implementation based on spin spirals} \\
The OAT dynamics is implemented via spin spirals using the method of Ref.~\cite{put_collective_2025}, with the full control sequence plotted as ``$k^\mathrm{th}$ stage OAT" in Fig.~\ref{fig2}a. In this control sequence, we first wind a spin spiral with wavevector $Q$, achieved through free evolution in a linear magnetic-field gradient. The winding is divided into two equal-duration segments separated by a MW $\pi$-pulse, which serves to refocus strong on-site disorder present in the sample. The resulting spiral then evolves under an XY8-type Floquet sequence that robustly decouples on-site disorder while preserving the form of the spin-spin interactions. The duration of the many-body quench is controlled by varying the number of Floquet periods. Finally, the spin spiral is unwound by another pair of gradient pulses (corresponding wavevector $-Q$) to refocus the spins.
\\

\noindent\textbf{Gradient characterization and control parameters} \\
The magnetic field generated by each of the four microwires is characterized individually under DC current using the NV ensemble as a vector magnetometer. Specifically, we measure all eight peaks of the electron spin resonance (ESR) spectrum and fit the resonance frequencies to extract the local vector magnetic field. The magnetic-field gradient is then obtained by repeating this measurement at five positions across the wire gap and performing a quadratic fit to the spatial dependence. The characterized fields and gradients produce by each wire are subsequently used to determine a current combination that generate a magnetic-field gradient aligned with the quantization axis ($\hat{Q}\parallel\hat{B_\perp}$) while maintaining zero net magnetic field at the centre of the ensemble. This gradient orientation is chosen to maximize the collective twisting rate~\cite{put_collective_2025}.

For spiral winding at a given wavevector $Q$, there is flexibility in the choice of gradient strength, which in turn determines the required winding time. For the wavevector $Q/2\pi = 3.3~$\textmu m$^{-1}$ used to generate the OAT dynamics, we chose a gradient strength of $\nabla B = 10~\mathrm{G}/$\textmu m in combination with a winding time of $2\times52~\mathrm{ns}$ (total winding time on both sides of the $\pi$-pulse). This gradient strength represents a trade-off between shorter winding time (favors stronger gradients) and reduced shot-to-shot variations in the gradient pulse area (favor weaker gradients, see Extended Data Fig.~\ref{ED_fig4}). 

To prevent transient gradient fields from impacting the MW drive, an additional padding time (35~ns) per rise/fall edge of the gradient is applied between the gradient pulses and neighboring MW pulse. A calibrated additional relative delay of 60~ns is applied to compensate for electrical delays in the circuitry.

For the multi-stage amplification protocols shown in Fig.~\ref{fig2} and Fig.~\ref{fig3}, we alternate the OAT-generating wavevector between $+Q$ and $-Q$ in odd/even amplification stages. This alternation helps preserve the Bloch-vector length during repeated amplification stages (Extended Data Fig.~\ref{ED_fig5}).
\\

\noindent\textbf{Floquet pulse sequence}\\
The disorder-decoupling pulse sequence used in this work is a modified XY8 sequence in which each $\pi$-pulse is replaced by a composite Knill $\pi$-pulse. Each Knill pulse consists of five simple $\pi$ pulses with different phases~\cite{ryan_robust_2010}. This sequence enables a modest improvement in coherence time compared to the ``cXY8'' sequence used in Ref.~\cite{gao_dressed-state_2026} (Extended Data Fig.~\ref{ED_fig6}). We note that this sequence is closely related to the KDD sequence introduced in Ref.~\cite{souza_robust_2011}, which is a modified XY4 sequence employing the same composite-pulse replacement.

The duration of each simple $\pi$-pulse is set to $t_\pi=36~$ns, and the free-evolution time between neighboring pulses is $t_\mathrm{free}=4~$ns. A cosine pulse envelope is used for each pulse to suppress off-resonant driving of the undesired $\ket{\tilde{0}}\leftrightarrow\ket{D}$ transition~\cite{gao_dressed-state_2026}.
\\

\noindent\textbf{Sensitivity calculation}\\
The sensitivities reported in Fig.~\ref{fig3}c are extracted from measured response curves (Fig.~\ref{fig3}b) according to the equation
\begin{equation}
    \mathrm{S}=\frac{\sigma_C}{\mathrm{d}C/\mathrm{d}B_\mathrm{AC}}\sqrt{NT_\mathrm{cycle}}.
\label{eq:sensitivity_definition}
\end{equation}
Here $C$ is the measured optical contrast with uncertainty $\sigma_C$ (see Extended Data Fig.~\ref{ED_fig2.5}), $\mathrm{d}C/\mathrm{d}B_\mathrm{AC}$ is the linear response near zero field, $N$ is the number of experimental repetitions, and $T_\mathrm{cycle}$ is the full cycle time, including phase accumulation, spiral winding and unwinding, and a fixed time overhead for spin initialization and readout. All experimental overheads associated to the interacting protocol are included in $T_\mathrm{cycle}$, providing a practical comparison with non-interacting protocols.
\\

\noindent\textbf{Numerical models and protocol optimizations}\\
In this work, we used a system model to generate the predictions shown in Fig.~\ref{fig2}c, and a phenomenological model for the predictions in Fig.~\ref{fig3}c,e. The phenomenological model is also used to optimize parameters in sensing protocols, including the durations of the initial sensing and OAT stages, as well as the turn angles.

The system model is based on discrete truncated Wigner approximation (DTWA)~\cite{schachenmayer_many-body_2015}.
A spin ensemble with density $\rho = 8.66\times10^{-5}~\mathrm{nm}^{-3}$ is simulated within a cylindrical geometry (diameter $1~$\textmu m, thickness $185~\mathrm{nm}$), with a smooth spin-polarization profile following a saturated optical-pumping model. The maximum spin polarization at the centre of the ensemble is assumed to be 80\%. In addition to the coherent dynamics governed by the dipolar interaction in Eq.~(\ref{eq:Heisenberg_interaction}), we include an experimentally characterized extrinsic coherence decay with $T_2 = 208~$\textmu s, modelled as probabilistic spin flips.

The phenomenological model tracks the collective spin polarization and incorporates experimentally measured parameters, including the twisting rate ($\mathrm{J_{OAT}}=99~\mathrm{kHz}$) and decay rate ($\gamma = 8.9~\mathrm{kHz}$) for a spiral wavevector $Q/2\pi = 3.3~$ \textmu m$^{-1}$, as well as an extrinsic coherence decay ($T_2 = 208~$\textmu s) when the spiral is unwound. For each winding-unwinding pair (\textit{i.e.} each OAT stage), we additionally include a coherence loss ($\sim 5\%$), and a global phase noise (std value $0.09~\mathrm{rad}$) that can be amplified in subsequent OAT stages (Extended Data Fig.~\ref{ED_fig4}). This global phase noise originates from shot-to-shot variations in the gradient pulses.

Signal and noise sensing protocols are numerically optimized using this phenomenological model. Three parameters are optimized: the duration of the initial sensing stage $t_\mathrm{sense}$, the duration of each OAT stage $t_\mathrm{OAT}$, and the target signal orientation immediately after each turn pulse $\theta_\mathrm{ZY}$ (defined as the angle between the Z axis and the red–blue separation in Fig.~\ref{fig2}d). For simplicity, the protocols are restricted to fixed values of $t_\mathrm{OAT}$ and $\theta_\mathrm{ZY}$ across all OAT stages. For coherent signal sensing (Fig.~\ref{fig3}b,c), the optimization is performed under the constraint of a fixed total phase-accumulation time $t_\mathrm{sense}+n t_\mathrm{OAT}$. For noise sensing (Fig.~\ref{fig3}d,e), all three parameters are optimized independently for each noise bandwidth. The noise sensitivity is evaluated by randomly sampling 50,000 noise realizations for each bandwidth. The optimized protocol parameters for signal sensing and noise sensing are summarized in Extended Data Fig.~\ref{ED_fig7} and Extended Data Fig.~\ref{ED_fig8}, respectively.
\\

\noindent\textbf{Experimental calibrations}\\
The experiments reported here require two key experimental calibrations: the spiral winding and unwinding times, and the phase and amplitude of the MW turn pulses.

Due to transient memory effects in the pulser circuitry, the pulse area of a gradient pulse can be weakly influenced by preceding gradient pulses when they occur in close temporal proximity (for example, when no Floquet period separates successive gradient pulses). As a result, the effective winding and unwinding times must be experimentally calibrated in these cases.

If the gradient pulse to be calibrated is an unwinding pulse (corresponding to the $-k_n$ pulse in Fig.~\ref{fig4}a), we sweep the unwinding time and measure the transverse (XY) coherence immediately after it. The measured coherence is fitted with a Gaussian function to find the maxima, which indicates optimal unwinding. If the pulse to be calibrated is a winding pulse (as in multi-stage OAT experiments and for the $K$ and $+Q$ pulses in Fig.~\ref{fig4}a) with a target effective duration $t_\mathrm{eff}$, we artificially add two Floquet periods after the target winding pulse, followed by an artificially added unwinding pulse with duration $t_\mathrm{eff}$. We then sweep the duration of the target winding pulse around $t_\mathrm{eff}$, and measure the transverse coherence after the added unwinding pulse to determine the optimal winding time. This procedure relies on the assumption that the added unwinding pulse is reliable, as it is well separated in time from preceding gradient pulses.

In multi-stage OAT experiments, the collective spin state can deviate from the X axis at the end of each OAT stage, even when the sensing signal is turned off. This deviation arises from accumulated pulse errors and pulser memory effects. To correct for this, we measure the phase of the transverse (XY) coherence immediately before the MW turn pulse (under zero sensing signal), and adjust the phase of the turn pulse accordingly. In addition, we measure the orientation of the sensing signal immediately before the turn pulse (defined as the angle between the Z axis and the red–blue separation in Fig.~\ref{fig2}d), and calculate the required turn angle by subtracting this measured value from the protocol parameter $\theta_\mathrm{ZY}$.
\\

\noindent\textbf{Data Availability}
The data that support the findings of this study are available from the corresponding author on request.
\\

\noindent\textbf{Code Availability}
The code used for data analysis and numerical modeling in this study is available from the corresponding author on request.
\\

\noindent\textbf{Acknowledgements}
We thank James MacArthur for technical contributions on this work. This work was supported by the National Science Foundation (grant number PHY-2012023), the Center for Ultracold Atoms (an NSF Physics Frontiers Center), Gordon and Betty Moore Foundation Grant No. 7797-01, the U.S. Department of Energy [DOE Quantum Systems Accelerator Center (Contract No.: DE-AC02-05CH11231) and BES grant No. DE-SC0019241], and the Army Research Office through the MURI program grant number W911NF-20-1-0136. We acknowledge the use of shared facilities of the UCSB Quantum Foundry through Q-AMASE-i program (NSF DMR-1906325), the UCSB MRSEC (NSF DMR 1720256), and the Quantum Structures Facility within the UCSB California NanoSystems Institute. 
A.C.B.J. acknowledge support from the NSF QLCI program through grant number OMA-2016245. 
L.B.H.W. acknowledges support from the NSF Graduate Research Fellowship Program (DGE 2139319) and the UCSB Quantum Foundry.
\\

\noindent\textbf{Author contributions} N.T.L. conceived the mechanism of collective OAT dynamics and proposed its interaction-enhanced sensing applications. H.G. and P.P. designed the experiments and analyzed the data. N.U.K. and N.T.L. developed the correlation sensing protocol. A.M. designed the gradient chip and the interfacing PCB. M.M. fabricated the gradient chip. H.G., N.T.L., and S.D  developed the dressed-state qubit encoding. L.B.H.W. fabricated the diamond sample. L.S.M. contributed to early-stage planning of the project. A.C.B.J., H.P., and M.D.L. supervised the project. All authors discussed the results and contributed to the manuscript.
\\

\noindent\textbf{Competing interests:} The authors declare no competing interests.
\\

\noindent\textbf{Additional Information:} 
Correspondence and requests for materials should be addressed to M.D.L.
\\

\setcounter{figure}{0}
\newcounter{EDfig}
\renewcommand{\figurename}{Extended Data Fig.}

\begin{figure*}
\centering
\includegraphics[width=2.0\columnwidth]{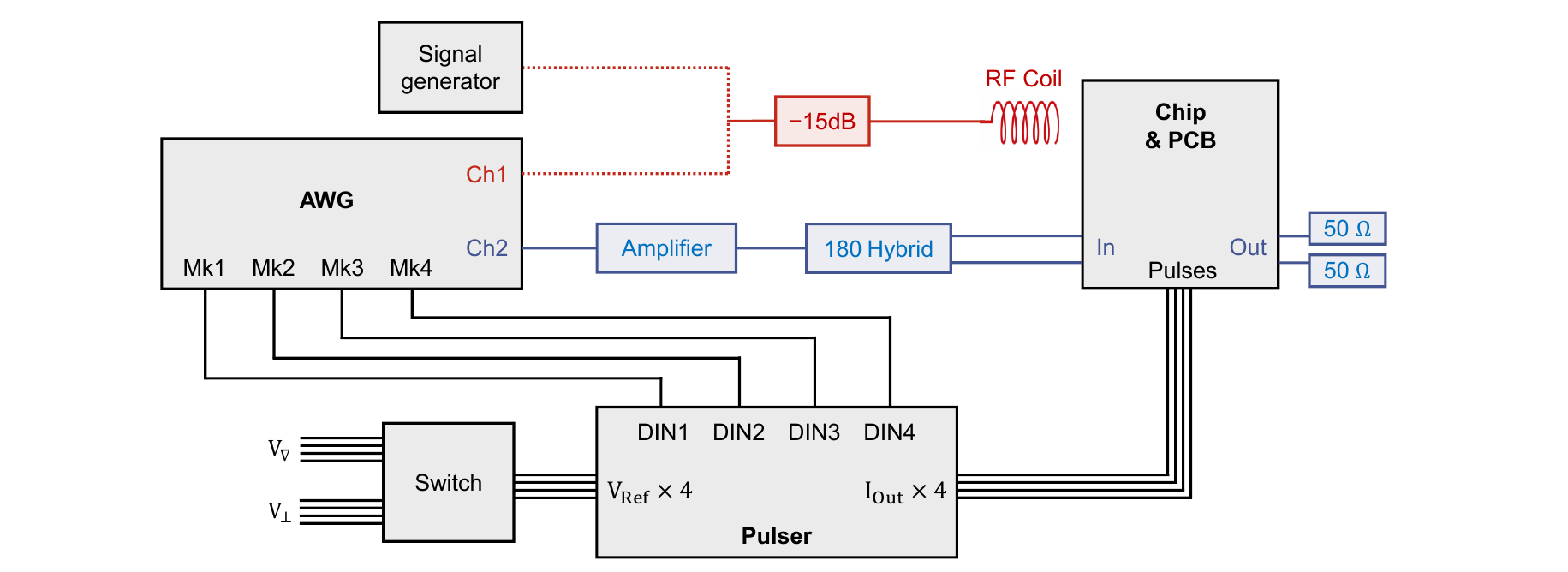}
\caption{\textbf{Schematic of the control circuitry.} The MW drive is generated by an analog output channel of the AWG, with relevant components indicated in blue. The target sensing signal and noise, indicated in red, are generated by another analog output channel of the AWG and a separate signal generator, respectively. The current pulser is controlled by the four marker channels of the AWG: Mk1 and Mk3 control the on/off state of pulses applied to two pairs of microwires, while Mk2 and Mk4 control the current polarity. The current amplitudes on the four microwires are set by four reference voltages $V_\mathrm{Ref}$, which are switched between two sets of values to alternate between the gradient and perpendicular-field configurations. The MW drive and current pulses are diplexed on the interfacing PCB and delivered to the chip (Extended Data Fig.~\ref{ED_fig2}). The AWG is Tektronix 7122C, the signal generator is Agilent 33522A, the MW amplifier is Mini-Circuits ZHL-16W-43-S+, the $180^\circ$ hybrid coupler is Fairview Microwave FMCP1155, and the switch is Analog Devices ADG1534.}
\refstepcounter{EDfig}\label{ED_fig1}
\end{figure*}

\begin{figure*}
\centering
\includegraphics[width=2.0\columnwidth]{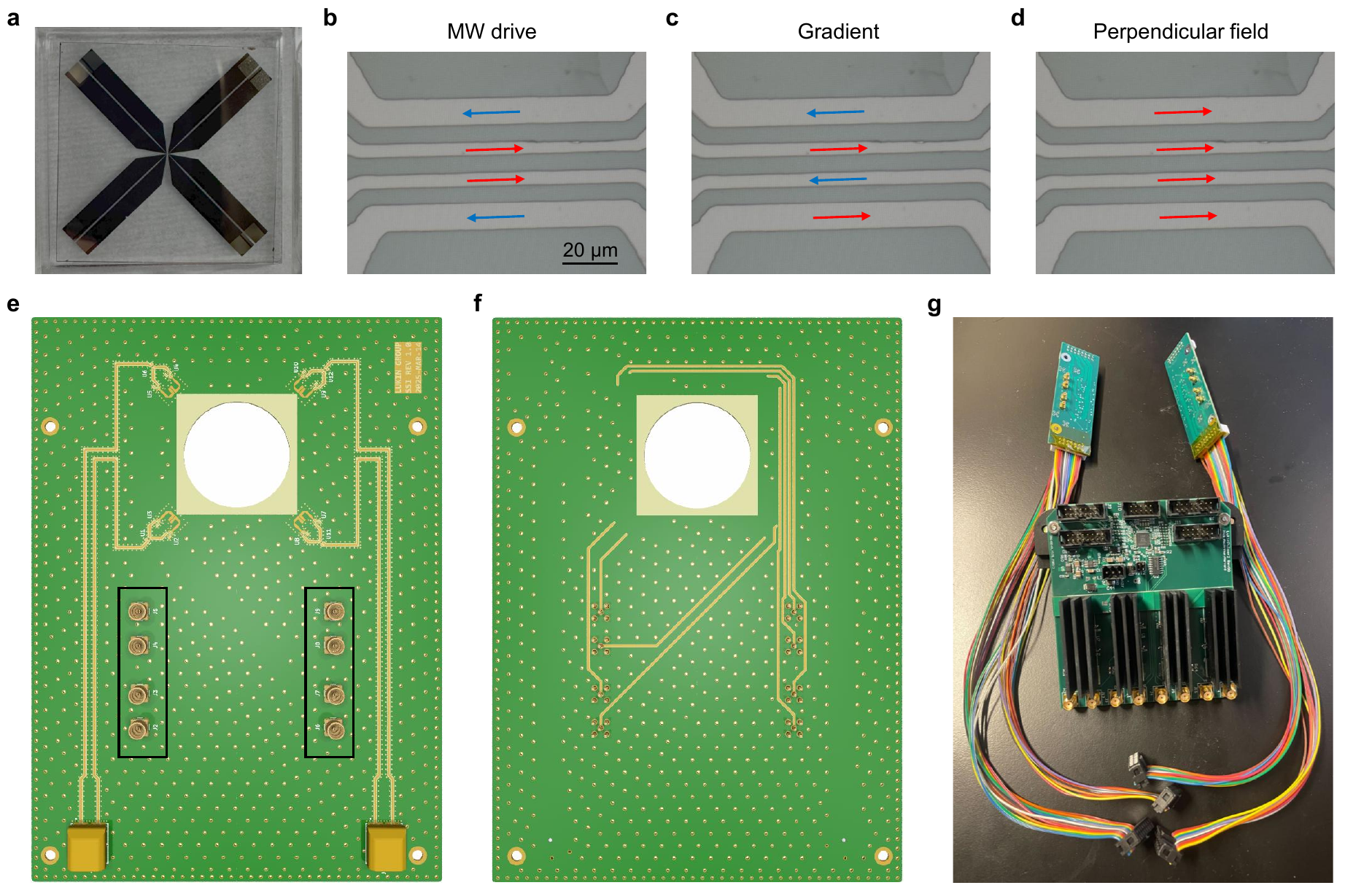}
\caption{\textbf{Custom-built circuit components.}  
\textbf{a,} Photo of the gradient chip. The reflective pads at the ends of the wires are bonding pads used for wire bonding to the interfacing PCB.
\textbf{b-d,} Microscope images of the central region of the gradient chip, with arrows indicating the current polarities for the three control fields.
\textbf{e,f,} Front and back sides of the interfacing PCB. The square near the top is reserved for the gradient chip, with the circular hole providing optical access. Two MW signals with a relative phase shift of $180^\circ$ are fed into a pair of connectors at the bottom of the PCB. These signals propagate through baluns, where they are converted into two pairs of differential signals, producing the target current polarity in panel b. After passing through the gradient chip, the differential signals are recombined by baluns and terminated at the remaining connectors at the bottom. The connectors in the black rectangles are used for interfacing with the current pulser (panel g). Pulsed currents are routed on the back side of the PCB and are diplexed with the MW drive after the baluns.
\textbf{g,} Photo of the current pulser. The pulser comprises a main board and two daughter boards, following the design described in Ref.~\cite{put_collective_2025}.
}
\refstepcounter{EDfig}\label{ED_fig2}
\end{figure*}

\begin{figure*}
\centering
\includegraphics[width=2.0\columnwidth]{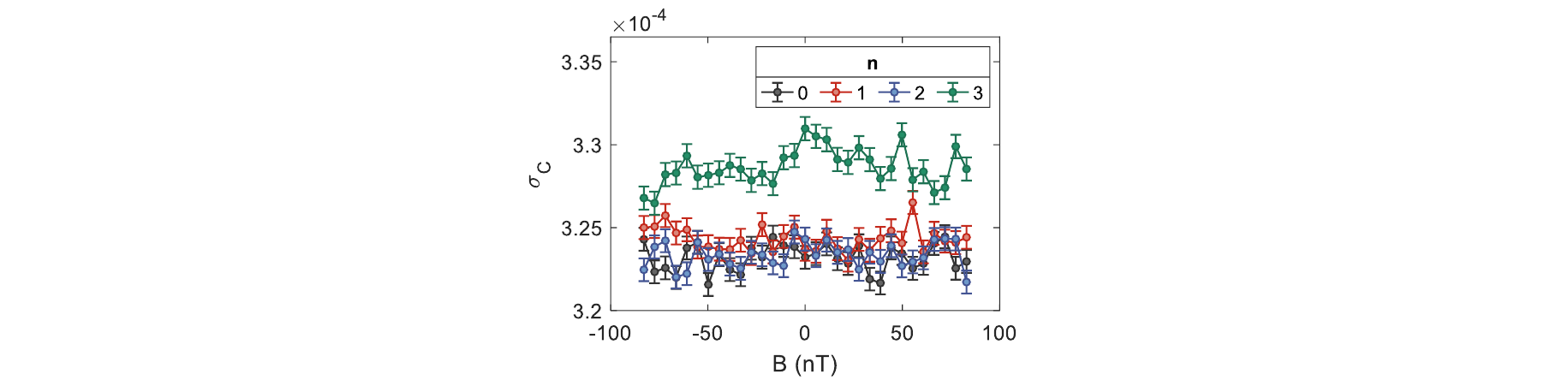}
\caption{\textbf{Behavior of optical contrast noise.}  
The uncertainty of measured optical contrast ($\sigma_C$ in Eq.~(\ref{eq:sensitivity_definition})) corresponding to the data in Fig.~\ref{fig3}b. The noise strength does not change substantially between interacting and non-interacting protocols (note the vertical scale), because the total readout noise is dominated by photon shot noise~\cite{shields_efficient_2015}, which is not affected by many-body dynamics. The $n=3$ measurement shows a slightly higher noise level. We believe that the increase in noise is most likely due to laser power drift. Errors represent 1~s.d. accounting statistical uncertainties.
}
\refstepcounter{EDfig}\label{ED_fig2.5}
\end{figure*}

\begin{figure*}
\centering
\includegraphics[width=2.0\columnwidth]{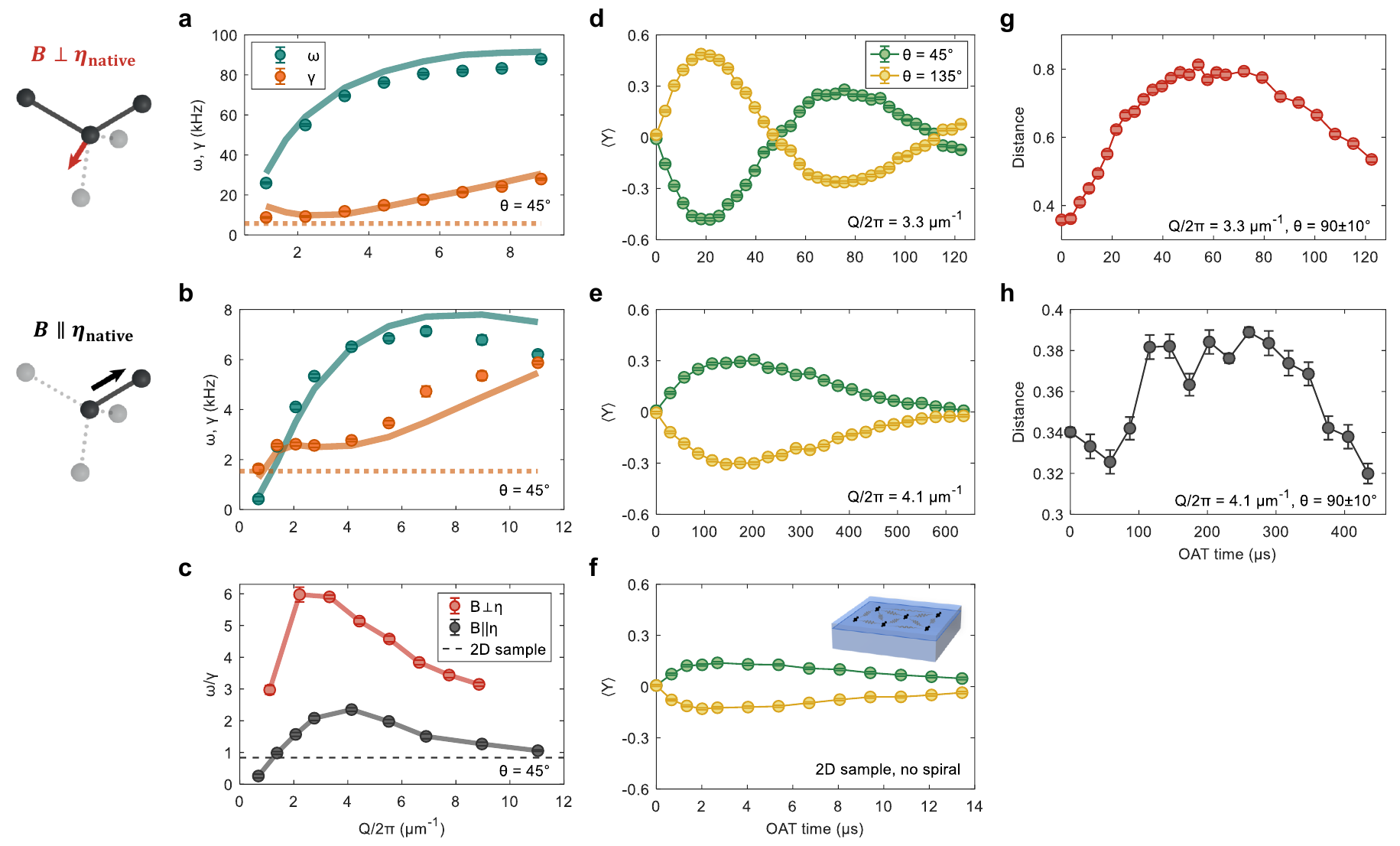}
\caption{\textbf{Comparison with previous works.}
First row: spiral-based OAT in the perpendicular-field qubit encoding (this work). Second row: spiral-based OAT in the aligned-field qubit encoding (Ref.~\cite{put_collective_2025}).
\textbf{a,b,} Measured twisting rate $\omega$ and transverse decay rate $\gamma$ as a function of spiral wavevector $Q$. The substantially larger twisting rate in the perpendicular-field encoding originates from the enhanced dipolar interaction strength $J$ (see Ref.~\cite{gao_dressed-state_2026}), while the higher OAT quality factor $\omega/\gamma$ arises from the improved dimensionless coherence time $JT_2$.
\textbf{c,} OAT quality factor $\omega/\gamma$ extracted from panels a and b. The dashed horizontal line corresponds to previous work~\cite{gao_signal_2025} based on native dipolar interactions in a 2D spin ensemble, where positional disorder corrupts collective dynamics and results in a lower OAT quality factor.
\textbf{d,e,} Measured Y polarization dynamics at the spiral wavevectors that maximize the OAT quality factor, shown for initial polar angles $\theta=45^\circ$ and $135^\circ$. Full oscillations are observed in the perpendicular-field encoding, indicating higher-quality collective dynamics. The opposite signs of the measured signals in the two encodings arise from the opposite overall sign of the dipolar interaction~\cite{gao_dressed-state_2026}. 
\textbf{f,} Same measurement as in panels d and e, but for the 2D ensemble work mentioned in panel c.
\textbf{g,h,} Single-stage signal amplification for initial sensing angles $\delta\theta=\pm10^\circ$. The significantly larger amplification observed in the perpendicular-field encoding originates from its higher OAT quality factor, combined with the second-order nature of OAT-based signal amplification~\cite{gao_signal_2025}. Data in panels b and e are from Ref.~\cite{put_collective_2025}, data in panel f are from Ref.~\cite{gao_signal_2025}, and data in panel a are from Fig.~\ref{fig2}c. Errors represent 1~s.d. accounting statistical uncertainties.
}
\refstepcounter{EDfig}\label{ED_fig3}
\end{figure*}

\begin{figure*}
\centering
\includegraphics[width=2.0\columnwidth]{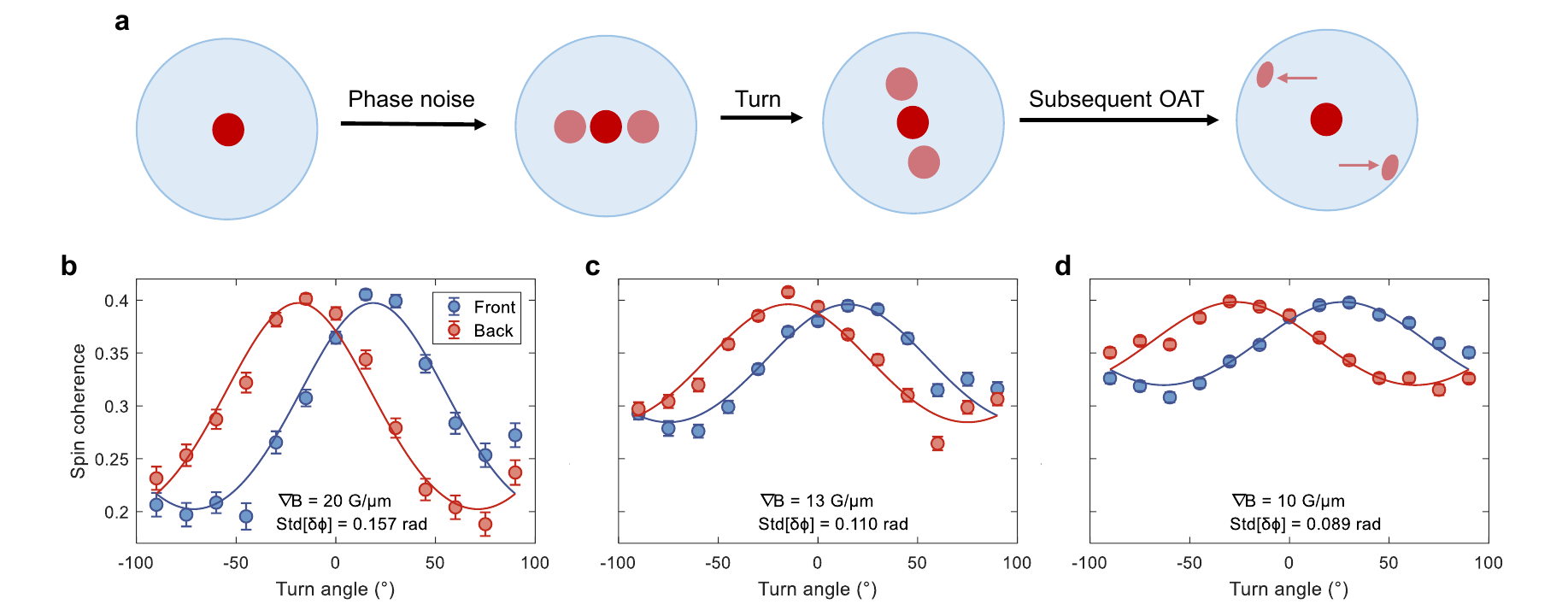}
\caption{\textbf{Phase-noise amplification in multi-stage OAT protocols.} \textbf{a,} The mechanism of phase-noise amplification. Phase noise is generated in each OAT stage due to shot-to-shot variations of the gradient pulses. The noise is rotated toward the Z axis by the turn pulse and subsequently amplified in later OAT stages. \textbf{b-d,} Measured transverse spin coherence following two-stage OAT protocols. The turn angle is swept while keeping the first- and second-stage durations fixed at $16~$\textmu s and $80~$\textmu s, respectively. Measurements are performed for three gradient strengths, $\nabla B=20,~13, ~10~$G/\textmu m, with a fixed spiral wavevector $Q/2\pi = 3.3~$\textmu$\mathrm{m}^{-1}$ realized by adjusting the winding durations. Blue and red data correspond to initial states prepared on the front and back sides of the Bloch sphere (\textit{i.e.} +X and -X states). The characteristic ``$\Lambda$"-shaped turn-angle dependence arises from phase-noise amplification, with larger amplitudes indicating stronger phase noise. Solid curves are fits to a model based on the mechanism shown in panel a (see Supplement), and the extracted phase-noise amplitudes Std[$\delta\phi$] are indicated in each panels. Errors represent 1~s.d. accounting statistical uncertainties.
}
\refstepcounter{EDfig}\label{ED_fig4}
\end{figure*}

\begin{figure*}
\centering
\includegraphics[width=2.0\columnwidth]{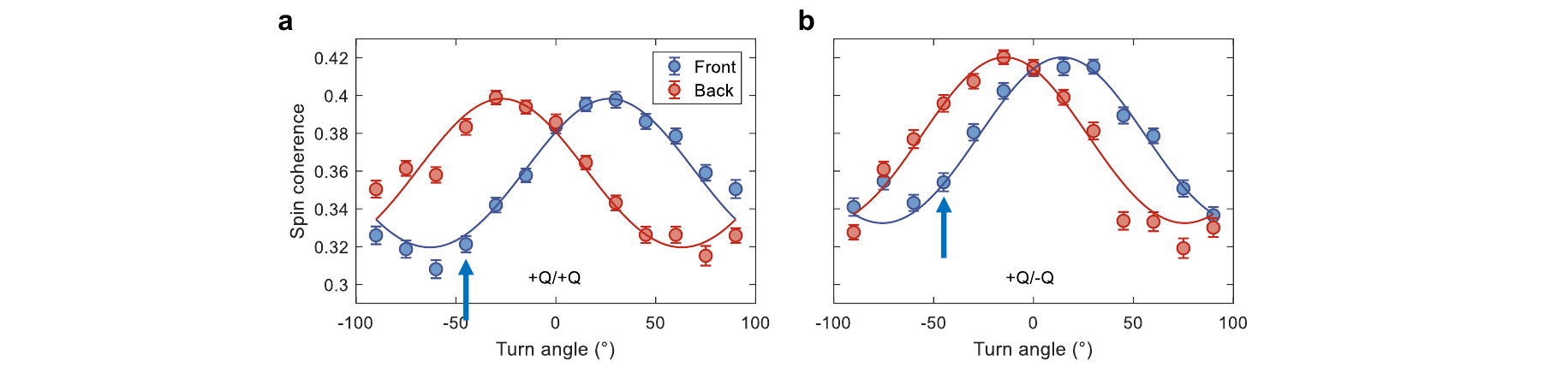}
\caption{\textbf{Motivation for alternating the OAT-generating wavevector.}  
\textbf{a,} Measured transverse spin coherence following a two-stage OAT protocol, with the OAT-generating wavevector fixed to $+Q$ in both stages. The turn angle is swept while keeping the first- and second-stage durations fixed at $16~$\textmu s and $80~$\textmu s, respectively. The blue arrow indicates a turn angle of $-45^\circ$, close to the value typically used in sensing experiments. \textbf{b,} Same measurement performed with the OAT-generating wavevector alternated between $+Q$ and $-Q$ in the two stages. This strategy results in a larger Bloch-vector length at the end of the protocol, indicating reduced coherence loss. See Supplement for explanations of this observation. Errors represent 1~s.d. accounting statistical uncertainties.
}
\refstepcounter{EDfig}\label{ED_fig5}
\end{figure*}

\begin{figure*}
\centering
\includegraphics[width=1.0\columnwidth]{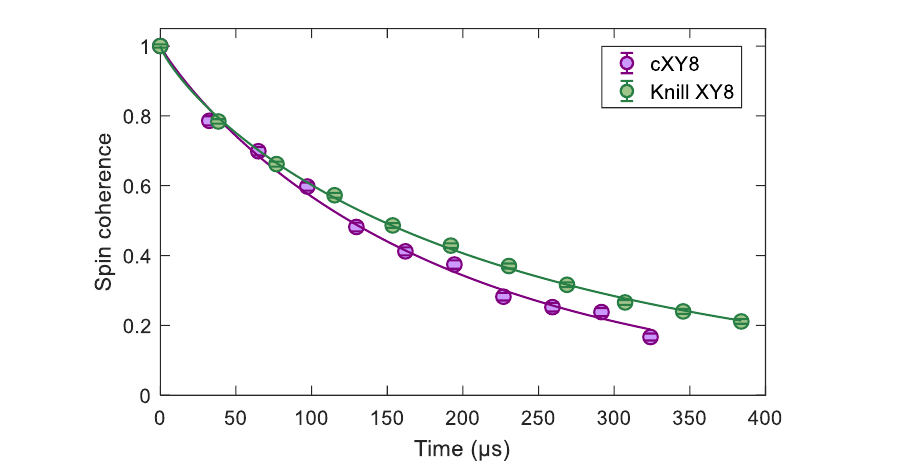}
\caption{\textbf{Coherence time comparison.} Measured transverse spin coherence as a function of evolution time for the ``Knill XY8" sequence (green) and the ``cXY8" sequence (purple). Solid curves are stretched-exponential fits to the data. ``Knill XY8" is used in this work, which slightly outperform ``cXY8" used in Ref.~\cite{gao_dressed-state_2026}. Errors represent 1~s.d. accounting statistical uncertainties.
}
\refstepcounter{EDfig}\label{ED_fig6}
\end{figure*}

\begin{figure*}
\centering
\includegraphics[width=2.0\columnwidth]{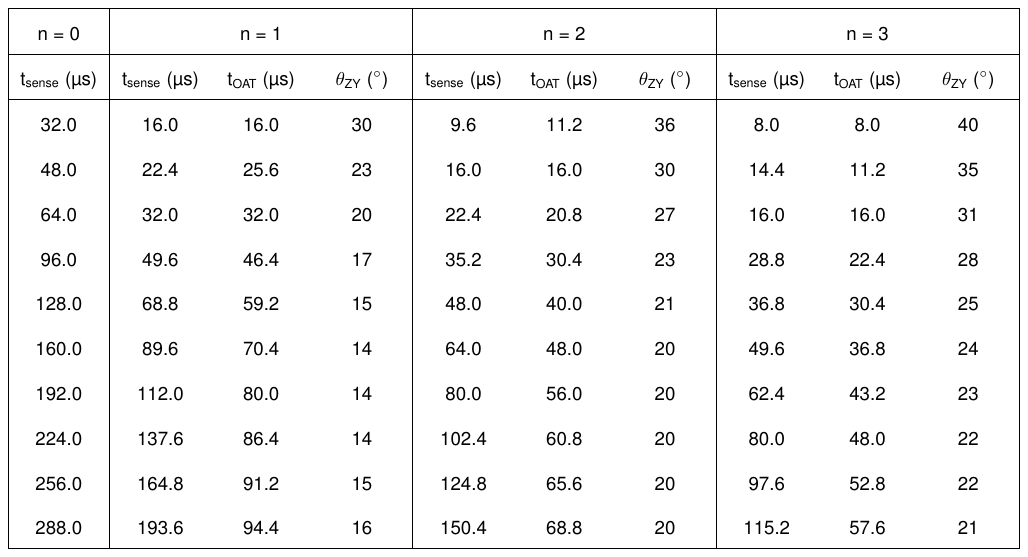}
\caption{\textbf{Protocol parameters for coherent signal sensing.} Numerically optimized protocol parameters used in the coherent-signal sensing experiments shown in Fig.~\ref{fig3}b,c, listed for each value of the total phase-accumulation time and for different numbers of amplification stages $n$. The parameters include the duration of the initial sensing stage $t_\mathrm{sense}$, the duration of each OAT stage $t_\mathrm{OAT}$, and the signal orientation immediately after each turn pulse $\theta_\mathrm{ZY}$ (defined as the angle between the Z axis and the red–blue separation in Fig.~\ref{fig2}d).
}
\refstepcounter{EDfig}\label{ED_fig7}
\end{figure*}

\begin{figure*}
\centering
\includegraphics[width=1.36\columnwidth]{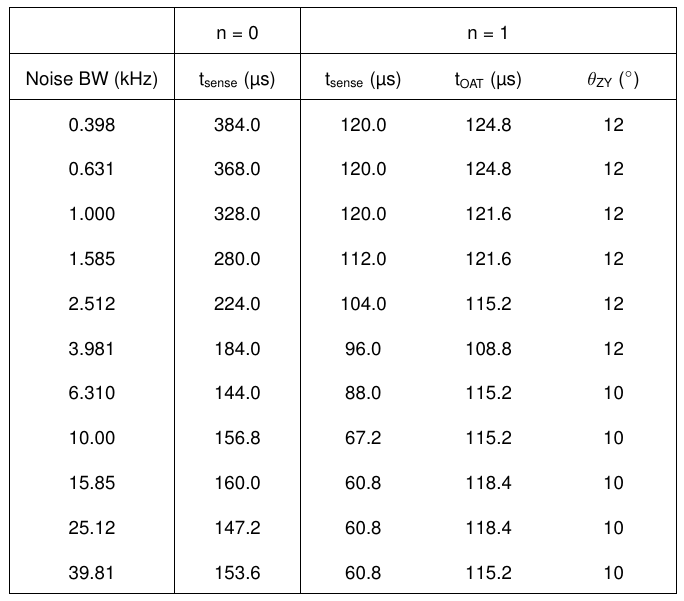}
\caption{\textbf{Protocol parameters for noise sensing.} Numerically optimized protocol parameters used in the noise sensing experiments shown in Fig.~\ref{fig3}d,e, listed for each value of the noise bandwidth and for different numbers of amplification stages $n$.
}
\refstepcounter{EDfig}\label{ED_fig8}
\end{figure*}

\begin{figure*}
\centering
\includegraphics[width=0.86\columnwidth]{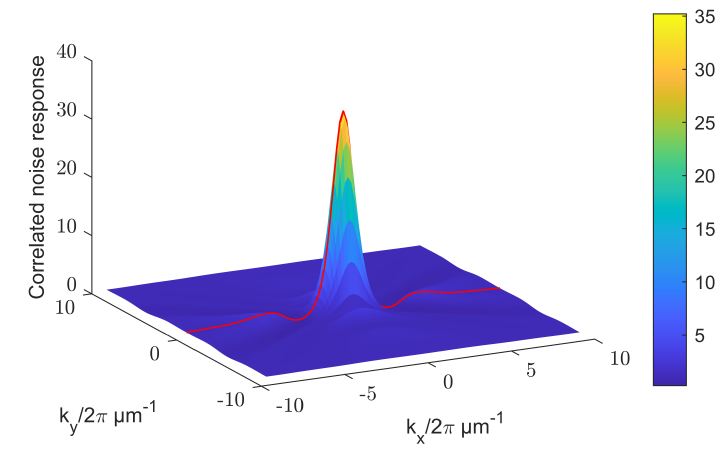}
\caption{\textbf{Predicted two-dimensional momentum-space response.} The response function $F(k)$ defined in Eq.~(\ref{eq:response_function}) is predicted as a function of the 2D wavevector $(k_x,~k_y)$, using a mean-field spin-wave model (see Supplement). $k_x$ and $k_y$ correspond to directions parallel and perpendicular to the OAT-generating wavevector $Q$, respectively. The red curve highlights the 1D cross-section along $k_x$, which is probed experimentally in Fig.~\ref{fig4}b.
}
\refstepcounter{EDfig}\label{ED_fig9}
\end{figure*}

\clearpage
\newpage

\onecolumngrid

\section*{Supplement Information}
\subsection{Phase-noise amplification model}
In this section, we describe the fitting model used for Extended Data Fig.~4b–d, based on the phase-noise amplification mechanism illustrated in Extended Data Fig.~4a.

We consider a two-stage OAT protocol. In the first stage, a phase noise with standard deviation $\delta\phi$ is generated. This phase noise is then rotated toward the Z axis by a turn pulse with angle $\alpha$, followed by amplification in the second OAT stage. We denote the duration of the second OAT stage by $t$, and the transverse spin coherence after the first OAT stage by $X_0$.

We now track the evolution of the spin state through each step shown in Extended Data Fig.~4a for a given realization of the phase noise $\delta\phi$. After the phase noise is introduced, the spin acquires a Y polarization of $(\delta\phi)X_0$. Following application of the turn pulse, the spin polarization is redistributed between Y- and Z-axes according to
\begin{align}
    \langle\mathrm{Y}\rangle &= (\delta\phi)X_0\mathrm{cos}\alpha,\nonumber\\
    \langle\mathrm{Z}\rangle &= (\delta\phi)X_0\mathrm{sin}\alpha.
\end{align}
The resulting Z polarization drives twisting dynamics during the second OAT stage, with twisting rate
\begin{equation}
    \omega = \mathrm{J_{OAT}}\langle\mathrm{Z}\rangle.
\end{equation}
This twisting rate produces an additional phase accumulation during the second OAT stage. Combining this contribution with the residual phase component remaining after the turn pulse ($(\delta\phi)\mathrm{cos}\alpha$) leads to a total accumulated phase at the end of the protocol,
\begin{equation}
    \phi_\mathrm{Final} = (\delta\phi)(\mathrm{cos}\alpha + \mathrm{J_{OAT}}t X_0\mathrm{sin}\alpha).
\end{equation}
In addition to this accumulated phase, the spin state acquires a finite latitude angle following the turn pulse ($\theta=(\delta\phi)\mathrm{sin}\alpha$), which also contributes to decoherence. Together, these effects lead to an additional decay of the transverse spin coherence at the end of the protocol, given by
\begin{align}
    D &= \mathrm{exp}\left\{-(\phi_\mathrm{Final}^2 + \theta^2)/2\right\}\nonumber\\
    &= \mathrm{exp}\left\{-\left[1+(\mathrm{J_{OAT}}t)^2X_0^2\mathrm{sin}^2\alpha + 2(\mathrm{J_{OAT}}t)X_0\mathrm{sin}\alpha\mathrm{cos}\alpha\right](\delta\phi)^2/2\right\}.
\end{align}

Including an overall spin coherence $A$ and a phenomenological phase offset $c$ that better captures the experimental data, we fitted the data in Extended Data Fig.~4b-d using the following fitting model:
\begin{equation}
    \langle\mathrm{X}\rangle = A\cdot\mathrm{exp}\left\{-\left[1+(\mathrm{J_{OAT}}t)^2X_0^2\mathrm{sin}^2(\alpha - c) + 2(\mathrm{J_{OAT}}t)X_0\mathrm{sin}(\alpha - c)\mathrm{cos}(\alpha - c)\right](\delta\phi)^2/2\right\}.
\end{equation}
Here $A$, $c$, and $\delta\phi$ are treated as fitting parameters, while all other quantities are independently measured experimentally. For initial states prepared on the back side of the Bloch sphere (red data in Extended Data Fig.~4b-d), the turn angle $\alpha$ is replaced by $-\alpha$. The data for the front and back initial states are fitted jointly using this model.

\subsection{Explanations for Extended Data Fig.~5}
To understand the difference between panel a and b of Extended Data Fig.~5, we transform the $\text{SU}(2)$-symmetric Hamiltonian Eq.~(1) in the lab frame to the spiral frame associated with the OAT-generating wavevector $Q$:
\begin{equation}
    H\propto\sum_{ij}J(\vec{r}_{ij})[S_i^z S_j^z + \mathrm{cos}(\vec{Q}\cdot\vec{r}_{ij})(S_i^x S_j^x + S_i^y S_j^y) + \mathrm{sin}(\vec{Q}\cdot\vec{r}_{ij})(S_i^x S_j^y - S_i^y S_j^x)].
\label{eq:spiral_frame_H}
\end{equation}
In this expression, the first two terms — those proportional to $S_i^z S_j^z$ and to $\mathrm{cos}(\vec{Q}\cdot\vec{r}_{ij})(S_i^x S_j^x + S_i^y S_j^y)$ — give rise to the leading-order collective twisting dynamics~~\cite{leitao_scalable_2026}. The remaining term, proportional to $\mathrm{sin}(\vec{Q}\cdot\vec{r}_{ij})(S_i^x S_j^y - S_i^y S_j^x)$, is often neglected in coarse-grained models. This approximation is motivated by the fact that the sine factor is spatially antisymmetric, whereas the dominant $S^x$ spin polarization is spatially homogeneous, leading to strong cancellation under ideal conditions.

In the present system, however, the contribution of this term is not entirely negligible. The finite spatial extent of the polarized region, together with positional disorder in the spin ensemble, breaks exact spatial symmetry and allows this term to have a measurable effect. Importantly, this term is the only component in Eq.~(\ref{eq:spiral_frame_H}) that breaks the symmetry under $+Q\leftrightarrow-Q$. As a result, it is solely responsible for the observed difference between the fixed-$Q$ protocol shown in Extended Data Fig.~5a and the $Q$-alternating protocol shown in Extended Data Fig.~5b.

The improved performance of the $Q$-alternating protocol can be understood using a many-body echo intuition. By alternating between $+Q$ and $-Q$ in successive OAT stages, the contributions of the sine term are partially canceled. This cancellation reduces the accumulated decoherence due to this term and results in a larger Bloch-vector length at the end of the multi-stage protocol.

\subsection{Intuitions for $k$-selective amplification}
It is not surprising that a \emph{collective} OAT mechanism selectively amplifies \emph{globally correlated} magnetic noise. To provide more intuition on this, we compare two representative cases: noise at $k_n=0$ and $k_n=Q$, where $Q$ is the spiral wavevector used to generate the OAT dynamics. As in the main text, we consider a sensing protocol in which the spatially-varying noise is encoded as a Z-modulation on a coherent spin state — which can be achieved via a $\pi/2$ rotation after the phase-accumulation — followed by an OAT stage that attempts to amplify it.

\begin{figure*}
\centering
\includegraphics[width=0.4\columnwidth]{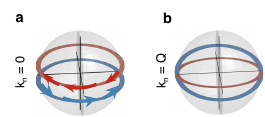}
\caption{\textbf{Comparison of two representative cases for $k$-selective amplification.}
\textbf{a,} Spin states immediately after spiral winding for the case of globally correlated noise (\textit{i.e.} $k_n=0$). Red and blue represent two noise instances with opposite signs, resulting in conical spin spirals on the upper and lower hemispheres. 
\textbf{b,} Same as panel a, but for the case of noise wavevector $k_n=Q$. This type of noise results in tilted spin spirals that lie along great circles of the Bloch sphere.}
\label{SI_Fig1}
\end{figure*}

The OAT stage begins with a spiral winding of wavevector $Q$. Immediately after the winding, noise at $k_n=0$ is mapped to a family of conical spin spirals, whose latitude on the Bloch sphere depends on the noise realization (Fig.~\ref{SI_Fig1}a). 
During the subsequent many-body quench, these conical spirals undergo twisting dynamics, resulting in an amplified response. In contrast, noise at $k_n=Q$ is mapped to tilted spin spirals that lie along great circles of the Bloch sphere (Fig.~\ref{SI_Fig1}b). This arises from the commensurate wavelengths of the noise-induced Z-modulation and the winding-induced transverse modulation. These spirals, equivalent to equatorial spin spirals ($\theta=\pi/2$) up to a global rotation, do not experience twisting dynamics and therefore do not lead to signal amplification. 

\subsection{Mean-field spin-wave model for correlation sensing}
In this section, we present the mean-field spin-wave model used for the fits in Fig.~4b,c, and for the prediction in Extended Data Fig.~9.

Sensing of a spatially correlated magnetic noise at wavevector $\vec{k}_n$ induces a spin modulation at the same wavevector. During the subsequent OAT stage, this modulation is encoded on top of a background spin spiral with wavevector $\vec{Q}$, and evolves under many-body dynamics, leading to wavevector-selective amplification. To quantitatively analyze this process, we transform to the spiral frame and analyze the dynamics starting from the spiral-frame Hamiltonian Eq.~(\ref{eq:spiral_frame_H}).

In the spiral frame, the unmodulated spin state is a coherent spin state along the X direction. A weak spin modulation with wavevector $k_n$ and phase $\phi_n$ yields
\begin{align}
    \langle S_i^x\rangle &= 1 - \mathcal{O}(\epsilon^2)\nonumber\\
    \langle S_i^y\rangle &= \epsilon_y \mathrm{cos}(\vec{k}_n\cdot\vec{r}_i+\phi_n)\nonumber\\
    \langle S_i^z\rangle &= \epsilon_z \mathrm{cos}(\vec{k}_n\cdot\vec{r}_i+\phi_n),
\label{eq:spin_modulation}
\end{align}
where $\epsilon_y$ and $\epsilon_z$ characterize the modulation amplitudes, and their dynamics under Eq.~(\ref{eq:spiral_frame_H}) determines the sensor response at wavevector $k_n$.

Here we focus on a mean-field analysis where the mean field
\begin{align}
    B_i^x &= \sum_j J(\vec{r}_{ij})\mathrm{cos}(\vec{Q}\cdot\vec{r}_{ij})\langle S_j^x\rangle + \sum_j J(\vec{r}_{ij})\mathrm{sin}(\vec{Q}\cdot\vec{r}_{ij})\langle S_j^y\rangle\nonumber\\
    B_i^y &= \sum_j J(\vec{r}_{ij})\mathrm{cos}(\vec{Q}\cdot\vec{r}_{ij})\langle S_j^y\rangle - \sum_j J(\vec{r}_{ij})\mathrm{sin}(\vec{Q}\cdot\vec{r}_{ij})\langle S_j^x\rangle\nonumber\\
    B_i^z &= \sum_j J(\vec{r}_{ij})\langle S_j^z\rangle
\label{eq:mean_field}
\end{align}
determines the spin evolution
\begin{align}
    \langle \dot S_i^y\rangle &= B_i^z \langle S_i^x\rangle - B_i^x \langle S_i^z\rangle\nonumber\\
    \langle \dot S_i^z\rangle &= B_i^x \langle S_i^y\rangle - B_i^y \langle S_i^x\rangle.
\label{eq:cross_product_evolution}
\end{align}
Plugging Eq.~(\ref{eq:spin_modulation},\ref{eq:mean_field}) into Eq.~(\ref{eq:cross_product_evolution}) and neglect higher order terms in $\epsilon_y$ and $\epsilon_z$, we get the dynamics of $\langle S_i^y\rangle$:
\begin{align}
    \langle \dot S_i^y\rangle &= \sum_j J(\vec{r}_{ij}) \langle S_j^z\rangle - \sum_j J(\vec{r}_{ij}) \mathrm{cos}(\vec{Q}\cdot\vec{r}_{ij})\langle S_i^z\rangle\nonumber\\
    &= \sum_j J(\vec{r}_{ij}) \epsilon_z\mathrm{cos}(\vec{k}_n\cdot\vec{r}_j + \phi_n)  - \sum_j J(\vec{r}_{ij}) \mathrm{cos}(\vec{Q}\cdot\vec{r}_{ij})\epsilon_z\mathrm{cos}(\vec{k}_n\cdot\vec{r}_i+\phi_n)\nonumber\\
    &= \sum_j J(\vec{r}_{ij}) \epsilon_z\left[\mathrm{cos}(\vec{k}_n\cdot\vec{r}_i + \phi_n)\mathrm{cos}(\vec{k}_n\cdot\vec{r}_{ij}) + \mathrm{sin}(\vec{k}_n\cdot\vec{r}_i + \phi_n)\mathrm{sin}(\vec{k}_n\cdot\vec{r}_{ij})\right]  - \sum_j J(\vec{r}_{ij}) \mathrm{cos}(\vec{Q}\cdot\vec{r}_{ij})\epsilon_z\mathrm{cos}(\vec{k}_n\cdot\vec{r}_i+\phi_n)\nonumber\\
    &= \sum_j J(\vec{r}_{ij}) \mathrm{cos}(\vec{k}_n\cdot\vec{r}_{ij})\epsilon_z\mathrm{cos}(\vec{k}_n\cdot\vec{r}_i + \phi_n)  - \sum_j J(\vec{r}_{ij}) \mathrm{cos}(\vec{Q}\cdot\vec{r}_{ij})\epsilon_z\mathrm{cos}(\vec{k}_n\cdot\vec{r}_i+\phi_n)\nonumber\\
    &= \sum_j J(\vec{r}_{ij}) \left[\mathrm{cos}(\vec{k}_n\cdot\vec{r}_{ij})-\mathrm{cos}(\vec{Q}\cdot\vec{r}_{ij})\right]\epsilon_z\mathrm{cos}(\vec{k}_n\cdot\vec{r}_i + \phi_n)\nonumber\\
    &= (\omega_{\vec{Q}} - \omega_{\vec{k}_n})\epsilon_z\mathrm{cos}(\vec{k}_n\cdot\vec{r}_i + \phi_n),
\label{eq:y_modulation_dynamics}
\end{align}
where $\omega_{\vec{Q}}$ is defined as
\begin{equation}
    \omega_{\vec{Q}}\equiv \sum_j J(\vec{r}_{ij})\left[1-\mathrm{cos}(\vec{Q}\cdot\vec{r}_{ij})\right].
\end{equation}
We note that this definition of $\omega_{\vec{Q}}$ is exactly the collective twisting rate associated with the OAT-generating wavevector $\vec{Q}$, as discussed in Refs.~~\cite{leitao_scalable_2026,put_collective_2025}. In the above derivation, the fourth line makes the approximation 
\begin{equation}
    \sum_j J(\vec{r}_{ij})\mathrm{sin}(\vec{k}_n\cdot\vec{r}_{ij})=0,
\label{eq:sine_approximation}
\end{equation}
which is justified by the spatial antisymmetry of the sine factor and the approximate spatial symmetry of the interaction kernel. An analogous analysis yields the dynamics of $\langle S_i^z\rangle$,
\begin{align}
    \langle \dot S_i^z\rangle &= \sum_j J(\vec{r}_{ij})\mathrm{cos}(\vec{Q}\cdot\vec{r}_{ij}) \langle S_i^y\rangle - \sum_j J(\vec{r}_{ij}) \mathrm{cos}(\vec{Q}\cdot\vec{r}_{ij})\langle S_j^y\rangle + \sum_j J(\vec{r}_{ij}) \mathrm{sin}(\vec{Q}\cdot\vec{r}_{ij})\nonumber\\
    &= \sum_j J(\vec{r}_{ij})\mathrm{cos}(\vec{Q}\cdot\vec{r}_{ij}) \epsilon_y\mathrm{cos}(\vec{k}_n\cdot\vec{r}_i+\phi_i) - \sum_j J(\vec{r}_{ij}) \mathrm{cos}(\vec{Q}\cdot\vec{r}_{ij})\epsilon_y\mathrm{cos}(\vec{k}_n\cdot\vec{r}_j+\phi_i)\nonumber\\
    &= \sum_j J(\vec{r}_{ij})\mathrm{cos}(\vec{Q}\cdot\vec{r}_{ij}) \epsilon_y\mathrm{cos}(\vec{k}_n\cdot\vec{r}_i+\phi_i) - \sum_j J(\vec{r}_{ij}) \mathrm{cos}(\vec{Q}\cdot\vec{r}_{ij})\epsilon_y\mathrm{cos}(\vec{k}_n\cdot\vec{r}_i+\phi_i)\mathrm{cos}(\vec{k}_n\cdot\vec{r}_{ij})\nonumber\\
    &= \sum_j J(\vec{r}_{ij})\left[\mathrm{cos}(\vec{Q}\cdot\vec{r}_{ij}) - \mathrm{cos}(\vec{Q}\cdot\vec{r}_{ij})\mathrm{cos}(\vec{k}_n\cdot\vec{r}_{ij})\right] \epsilon_y\mathrm{cos}(\vec{k}_n\cdot\vec{r}_i+\phi_i)\nonumber\\
    &= \left[\frac{\omega_{\vec{Q}+\vec{k}_n} + \omega_{\vec{Q}-\vec{k}_n}}{2} - \omega_{\vec{Q}}\right] \epsilon_y\mathrm{cos}(\vec{k}_n\cdot\vec{r}_i+\phi_i),
\label{eq:z_modulation_dynamics}
\end{align}
where the second line makes the approximation in Eq.~(\ref{eq:sine_approximation}) again, and the third line makes the approximation
\begin{equation}
    \sum_j J(\vec{r}_{ij})\mathrm{cos}(\vec{Q}\cdot\vec{r}_{ij})\mathrm{sin}(\vec{k}_n\cdot\vec{r}_{ij})=0,
\end{equation}
following a similar justification. Defining quantities $\delta_{\vec{k}_n}$ and $\Delta_{\vec{k}_n}$ as
\begin{align}
    \delta_{\vec{k}_n} &\equiv \omega_{\vec{Q}} - \omega_{\vec{k}_n}\nonumber\\
    \Delta_{\vec{k}_n} &\equiv \omega_{\vec{Q}}- \frac{\omega_{\vec{Q}+\vec{k}_n} + \omega_{\vec{Q}-\vec{k}_n}}{2},
\end{align}
Eq.~(\ref{eq:y_modulation_dynamics}, \ref{eq:z_modulation_dynamics}) yield the coupled equation of motion
\begin{align}
    \dot\epsilon_y &= \delta_{\vec{k}_n}\epsilon_z\nonumber\\
    \dot\epsilon_z &= -\Delta_{\vec{k}_n}\epsilon_y,
\label{eq:spin_wave_EOM}
\end{align}
which serve as the basis for further analysis.

\subsubsection{$\vec{K}=\vec{0}$ case, initial Z-modulation}
We first consider the case in which the mapping stage is absent and the initial spin modulation is along the Z direction. This case corresponds to the top panel of Fig.~4b, and to Extended Data Fig.~9.

The solution of Eq.~(\ref{eq:spin_wave_EOM}) is
\begin{align}
    \epsilon_z(t) &= \mathrm{cos}\left(\sqrt{\delta_{\vec{k}_n}\Delta_{\vec{k}_n}}t\right)\epsilon,\nonumber\\
    \epsilon_y(t) &= (\delta_{\vec{k}_n}t)\mathrm{sinc}\left(\sqrt{\delta_{\vec{k}_n}\Delta_{\vec{k}_n}}t\right)\epsilon,
\end{align}
with $\epsilon$ stands for the initial modulation amplitude $\epsilon_z(t=0)$. This solution corresponds to a noise amplification factor of
\begin{equation}
    A=\left[\epsilon_y^2(t) + \epsilon_z^2(t)\right]/\epsilon^2 = \mathrm{cos}^2\left(\sqrt{\delta_{\vec{k}_n}\Delta_{\vec{k}_n}}t\right) + (\delta_{\vec{k}_n}t)^2\mathrm{sinc}^2\left(\sqrt{\delta_{\vec{k}_n}\Delta_{\vec{k}_n}}t\right).
\label{eq:noise_amplify}
\end{equation}
In interaction-enhanced cosine magnetometry, the additional decay of the X polarization arises from amplified noise components along the 
Y and Z directions. As a result, the expression above directly determines the momentum-space response function $F(\vec{k}_n)$. To account for experimental imperfections and overall amplitude of the response, we introduce a fitting parameter $\beta$ that sets the effective amplitude of the response function. This yields the fitting model used for the data in the top panel of Fig.~4b,
\begin{equation}
    F_z(\vec{k}_n) = \mathrm{cos}^2\left(\beta\sqrt{\delta_{\vec{k}_n}\Delta_{\vec{k}_n}}t\right) + (\beta\delta_{\vec{k}_n}t)^2\mathrm{sinc}^2\left(\beta\sqrt{\delta_{\vec{k}_n}\Delta_{\vec{k}_n}}t\right).
\label{eq:F_z(k)}
\end{equation}
In the fitting, the quantities $\delta_{\vec{k}_n}$ and $\Delta_{\vec{k}_n}$ are extracted from the experimentally measured twisting rate $\omega_k$ reported in Fig.~2c, using cubic-spline-based interpolation. The fitted value of $\beta$ is used to generate the prediction in Extended Data Fig.~9.

For the experimentally probed configuration $\vec{k}_n\parallel\vec{Q}$, the expression above can be further simplified. In this geometry, $\delta_{\vec{k}_n}$ is peaked around $\vec{k}_n=\vec{0}$, whereas $\Delta_{\vec{k}_n}$ is peaked around $\vec{k}_n=\pm\vec{Q}$. Consequently, these two quantities are not simultaneously large, allowing the approximation $\sqrt{\delta_{\vec{k}_n}\Delta_{\vec{k}_n}}t \ll 1$. This leads to the simplified (but approximate) expression
\begin{equation}
    F_z(k_n) \approx 1+\beta^2(\omega_Q - \omega_{k_n})^2t^2.
\label{eq:F_z(k)_simplified}
\end{equation}
We emphasize that this approximation is valid only for the case $\vec{k}_n\parallel\vec{Q}$.

This simplified expression provides direct insight into the width of the response-function peak. Because the peak is governed by the difference in twisting rates $\omega_Q - \omega_{k_n}$, and the twisting rate saturates at a scale set by the inverse system size, approximately $1/L$~~\cite{leitao_scalable_2026,put_collective_2025}, the width of the response-function peak is likewise determined by $1/L$.

\subsubsection{$\vec{K}=\vec{0}$ case, initial Y-modulation}
We next consider the case in which the initial spin modulation is along the Y direction. Although this configuration is not directly probed in the experiments, its analysis is essential for understanding the $\vec{K}\neq\vec{0}$ cases, where the initial Z-modulation is mapped to both Y- and Z- modulations during the mapping stage.

The dynamics in this case differ from the initial Z-modulation case only by an exchange of the roles of the modulation amplitudes $\epsilon_y$ and $\epsilon_z$. As a result, the corresponding response function can be obtained directly by interchanging the quantities $\delta_{\vec{k}_n}$ and $\Delta_{\vec{k}_n}$ in Eq.~(\ref{eq:F_z(k)}). This yields
\begin{equation}
    F_y(\vec{k}_n) = \mathrm{cos}^2\left(\beta\sqrt{\delta_{\vec{k}_n}\Delta_{\vec{k}_n}}t\right) + (\beta\Delta_{\vec{k}_n}t)^2\mathrm{sinc}^2\left(\beta\sqrt{\delta_{\vec{k}_n}\Delta_{\vec{k}_n}}t\right),
\label{eq:F_y(k)}
\end{equation}
where $\beta$ is the fitting parameter introduced previously, with theoretical value $\beta = 1$. This response function is peaked around $\vec{k}_n=\pm \vec{Q}$, with smaller peak amplitude than $F_z(\vec{k}_n)$.

\subsubsection{$\vec{K}\neq\vec{0}$ case, initial Z-modulation}
Finally, we study the case with the mapping wavevector $\vec{K}\neq\vec{0}$, corresponding to the bottom two panels of Fig.~4b and to Fig.~4c.

We start with an initial Z-modulation
\begin{equation}
    \langle S_i^z\rangle = \epsilon\cdot\mathrm{cos}(\vec{k}_n\cdot\vec{r}_i+\phi_n),
\end{equation}
which is mapped by the mapping stage to both Y- and Z- modulations:
\begin{align}
    \langle S_i^y\rangle &= -\epsilon\cdot\mathrm{cos}(\vec{k}_n\cdot\vec{r}_i+\phi_n)\mathrm{sin}(\vec{K}\cdot\vec{r}_i)\nonumber\\
    &= \frac{\epsilon}{2}\mathrm{sin}\left[(\vec{k}_n-\vec{K})\cdot\vec{r}_i+\phi_n\right] - \frac{\epsilon}{2}\mathrm{sin}\left[(\vec{k}_n+\vec{K})\cdot\vec{r}_i+\phi_n\right]\nonumber\\
    \langle S_i^z\rangle &= \epsilon\cdot\mathrm{cos}(\vec{k}_n\cdot\vec{r}_i+\phi_n)\mathrm{cos}(\vec{K}\cdot\vec{r}_i)\nonumber\\
    &= \frac{\epsilon}{2}\mathrm{cos}\left[(\vec{k}_n-\vec{K})\cdot\vec{r}_i+\phi_n\right] + \frac{\epsilon}{2}\mathrm{cos}\left[(\vec{k}_n+\vec{K})\cdot\vec{r}_i+\phi_n\right].
\end{align}
The resulting spin modulation contains modes at both $\vec{k}_n+\vec{K}$ and $\vec{k}_n-\vec{K}$  which, to the leading order, evolve independently in time. The evolution of these modes follows the equation of motion Eq.~(\ref{eq:spin_wave_EOM}). At time $t$, they evolve to
\begin{align}
    \langle S_i^y\rangle &= \frac{\epsilon}{2}\mathrm{cos}(E_{\vec{k}_n-\vec{K}}t)\mathrm{sin}\left[(\vec{k}_n-\vec{K})\cdot\vec{r}_i + \phi_n\right] - \frac{\epsilon}{2}\mathrm{cos}(E_{\vec{k}_n+\vec{K}}t)\mathrm{sin}\left[(\vec{k}_n+\vec{K})\cdot\vec{r}_i + \phi_n\right]\nonumber\\
    &+ \frac{\epsilon}{2}(\delta_{\vec{k}_n-\vec{K}}t)\mathrm{sinc}(E_{\vec{k}_n-\vec{K}}t)\mathrm{cos}\left[(\vec{k}_n-\vec{K})\cdot\vec{r}_i + \phi_n\right] + \frac{\epsilon}{2}(\delta_{\vec{k}_n+\vec{K}}t)\mathrm{sinc}(E_{\vec{k}_n+\vec{K}}t)\mathrm{cos}\left[(\vec{k}_n+\vec{K})\cdot\vec{r}_i + \phi_n\right]\nonumber\\
    \langle S_i^z\rangle &= \frac{\epsilon}{2}\mathrm{cos}(E_{\vec{k}_n-\vec{K}}t)\mathrm{cos}\left[(\vec{k}_n-\vec{K})\cdot\vec{r}_i + \phi_n\right] + \frac{\epsilon}{2}\mathrm{cos}(E_{\vec{k}_n+\vec{K}}t)\mathrm{cos}\left[(\vec{k}_n+\vec{K})\cdot\vec{r}_i + \phi_n\right]\nonumber\\
    &- \frac{\epsilon}{2}(\Delta_{\vec{k}_n-\vec{K}}t)\mathrm{sinc}(E_{\vec{k}_n-\vec{K}}t)\mathrm{sin}\left[(\vec{k}_n-\vec{K})\cdot\vec{r}_i + \phi_n\right] + \frac{\epsilon}{2}(\Delta_{\vec{k}_n+\vec{K}}t)\mathrm{sinc}(E_{\vec{k}_n+\vec{K}}t)\mathrm{sin}\left[(\vec{k}_n+\vec{K})\cdot\vec{r}_i + \phi_n\right]
\end{align}
where $E_{\vec{k}}$ is defined as
\begin{equation}
    E_{\vec{k}} \equiv \sqrt{\delta_{\vec{k}}\Delta_{\vec{k}}}.
\end{equation}
To quantify the amplification of noise at each location, we define a local response function $F_i$
\begin{equation}
    F_i\equiv\frac{\langle S_i^y\rangle^2 + \langle S_i^z\rangle^2}{\epsilon^2/2},
\end{equation}
and compute its value while averaging over the noise phase $\phi_n$:
\begin{align}
    \overline{F_i} &= \frac{1}{2}\mathrm{cos}^2(E_{\vec{k}_n-\vec{K}}t) + \frac{1}{2}\mathrm{cos}^2(E_{\vec{k}_n+\vec{K}}t)\nonumber\\
    &+ \frac{1}{4}(\delta_{\vec{k}_n-\vec{K}}t)^2\mathrm{sinc}^2(E_{\vec{k}_n-\vec{K}}t) + \frac{1}{4}(\delta_{\vec{k}_n+\vec{K}}t)^2\mathrm{sinc}^2(E_{\vec{k}_n+\vec{K}}t)\nonumber\\
    &+ \frac{1}{4}(\Delta_{\vec{k}_n-\vec{K}}t)^2\mathrm{sinc}^2(E_{\vec{k}_n-\vec{K}}t) + \frac{1}{4}(\Delta_{\vec{k}_n+\vec{K}}t)^2\mathrm{sinc}^2(E_{\vec{k}_n+\vec{K}}t)\nonumber\\
    &+\frac{1}{2}\left[\delta_{\vec{k}_n-\vec{K}}\delta_{\vec{k}_n+\vec{K}} - \Delta_{\vec{k}_n-\vec{K}}\Delta_{\vec{k}_n+\vec{K}}\right]t^2\mathrm{sinc}(E_{\vec{k}_n-\vec{K}}t)\mathrm{sinc}(E_{\vec{k}_n+\vec{K}}t)\mathrm{cos}(2\vec{K}\cdot\vec{r}_i)\nonumber\\
    &+\frac{1}{2}\left[(\Delta_{\vec{k}_n-\vec{K}}-\delta_{\vec{k}_n-\vec{K}})\mathrm{sinc}(E_{\vec{k}_n-\vec{K}}t)\mathrm{cos}(E_{\vec{k}_n+\vec{K}}t) + (\Delta_{\vec{k}_n+\vec{K}}-\delta_{\vec{k}_n+\vec{K}})\mathrm{sinc}(E_{\vec{k}_n+\vec{K}}t)\mathrm{cos}(E_{\vec{k}_n-\vec{K}}t)\right]\mathrm{sin}(2\vec{K}\cdot\vec{r}_i)t.
\end{align}
The response function associated to the global readout can then be calculated by averaging over a Gaussian detection profile centered at $\vec{r}=\vec{0}$ with standard deviation $\sigma$, yielding
\begin{align}
    F_z(\vec{k}_n,\vec{K}) &= \frac{1}{2}\mathrm{cos}^2(E_{\vec{k}_n-\vec{K}}t) + \frac{1}{2}\mathrm{cos}^2(E_{\vec{k}_n+\vec{K}}t)\nonumber\\
    &+ \frac{1}{4}(\delta_{\vec{k}_n-\vec{K}}t)^2\mathrm{sinc}^2(E_{\vec{k}_n-\vec{K}}t) + \frac{1}{4}(\delta_{\vec{k}_n+\vec{K}}t)^2\mathrm{sinc}^2(E_{\vec{k}_n+\vec{K}}t)\nonumber\\
    &+ \frac{1}{4}(\Delta_{\vec{k}_n-\vec{K}}t)^2\mathrm{sinc}^2(E_{\vec{k}_n-\vec{K}}t) + \frac{1}{4}(\Delta_{\vec{k}_n+\vec{K}}t)^2\mathrm{sinc}^2(E_{\vec{k}_n+\vec{K}}t)\nonumber\\
    &+\frac{1}{2}\left[\delta_{\vec{k}_n-\vec{K}}\delta_{\vec{k}_n+\vec{K}} - \Delta_{\vec{k}_n-\vec{K}}\Delta_{\vec{k}_n+\vec{K}}\right]t^2\mathrm{sinc}(E_{\vec{k}_n-\vec{K}}t)\mathrm{sinc}(E_{\vec{k}_n+\vec{K}}t)\mathrm{e}^{-(2K\sigma)^2/2}.
\label{eq:response_function_K}
\end{align}
Recognizing the first three lines as $\vec{K}=\vec{0}$ response functions for initial Y- and Z-modulations, the above expression can be re-written into a compact form
\begin{align}
    F_z(\vec{k}_n,\vec{K}) &= \frac{1}{4}F_z(\vec{k}_n-\vec{K}) + \frac{1}{4}F_z(\vec{k}_n+\vec{K}) + \frac{1}{4}F_y(\vec{k}_n-\vec{K}) + \frac{1}{4}F_y(\vec{k}_n+\vec{K})\nonumber\\
    &+\frac{1}{2}\left[\delta_{\vec{k}_n-\vec{K}}\delta_{\vec{k}_n+\vec{K}} - \Delta_{\vec{k}_n-\vec{K}}\Delta_{\vec{k}_n+\vec{K}}\right]t^2\mathrm{sinc}(E_{\vec{k}_n-\vec{K}}t)\mathrm{sinc}(E_{\vec{k}_n+\vec{K}}t)\mathrm{e}^{-(2|K|\sigma)^2/2}.
\label{eq:response_function_K_short}
\end{align}

This form provides a transparent physical interpretation. At sufficiently large $|K|$, the exponential suppression eliminates the second line, and the response function reduces to a simple average of $F_y$ and $F_z$ at $\vec{k}_n\pm\vec{K}$, consistent with the intuition that the mapping stage redistribute the spin-modulation into Y- and Z-modulations at these two wavevectors. We note that the terms $F_y(\vec{k}_n-\vec{K})$ and $F_y(\vec{k}_n+\vec{K})$ are responsible for the side peaks observed in the bottom two panels of Fig.~4b, adjacent to the main response at $\vec{k}_n = \pm \vec{K}$.

Similar to before, we can introduce a fitting parameter $\beta$ to account for the overall amplitude. This parameter is used in fitting the data shown in the bottom two panels of Fig.~4b. For the data in Fig.~4c, an additional multiplicative factor of the form $\mathrm{e}^{-|K|/K_0}$ is included, with fitting parameter $K_0$, to capture decoherence incurred during the gradient-winding process, which increases for longer winding time (\textit{i.e.} larger $|K|$).

\end{document}